\documentclass[journal]{IEEEtran} 
\usepackage{CJKutf8}
\usepackage{cmap}
\usepackage[T1]{fontenc}
\usepackage{cite} 
\usepackage{amsmath,amssymb,amsfonts}
\usepackage{graphicx} 
\usepackage{textcomp}
\usepackage{xcolor}
\usepackage{hyperref}
\usepackage{bm}
\usepackage{amsthm}
\usepackage{enumerate}
\usepackage{booktabs} 
\usepackage{diagbox}
\usepackage{supertabular}
\usepackage{subfigure}
\usepackage{caption}
\usepackage{algorithmicx}
\usepackage[linesnumbered,ruled,vlined]{algorithm2e}

\begin{document} 
\begin{CJK}{UTF8}{gbsn}

\title{Graph Attention Network-based Block Propagation with Optimal AoB and Reputation in Web 3.0}

\author{
Jiana Liao*, Jinbo Wen*, Jiawen Kang, Changyan Yi, Yang Zhang, Yutao Jiao, \\Dusit Niyato, \textit{Fellow, IEEE}, Dong In Kim, \textit{Fellow, IEEE}, and Shengli Xie, \textit{Fellow, IEEE}

\thanks{
J. Liao, J. Kang, and S. Xie are with the School of Automation, Guangdong University of Technology, China (e-mails:  3221000946@mail2.gdut.edu.cn; kavinkang@gdut.edu.cn; shlxie@gdut.edu.cn).

J. Wen, C. Yi, and Y. Zhang are with the College of Computer Science and Technology, Nanjing University of Aeronautics and Astronautics, China (e-mails: jinbo1608@163.com; changyan.yi@nuaa.edu.cn; yangzhang@nuaa.edu.cn).

Y. Jiao is with the College of Communications Engineering, Army Engineering University of PLA, China (e-mail: yjiao001@yeah.net).

D. Niyato is with the School of Computer Science and Engineering, Nanyang Technological University, Singapore (e-mail: dniyato@ntu.edu.sg).

D. I. Kim is with the Department of Electrical and Computer Engineering, Sungkyunkwan University, Suwon 16419, South Korea (e-mail: dikim@skku.ac.kr). 

* means equal contribution. (\textit{Corresponding authors: Jiawen Kang, Yutao Jiao.})
}
}

\maketitle
\begin{abstract}
Web 3.0 is recognized as a pioneering paradigm that empowers users to securely oversee data without reliance on a centralized authority. Blockchains, as a core technology to realize Web 3.0, can facilitate decentralized and transparent data management. Nevertheless, the evolution of blockchain-enabled Web 3.0 is still in its nascent phase, grappling with challenges such as ensuring efficiency and reliability to enhance block propagation performance. In this paper, we design a Graph Attention Network (GAT)-based reliable block propagation optimization framework for blockchain-enabled Web 3.0. We first innovatively apply a data-freshness metric called age of block to measure block propagation efficiency in public blockchains. To achieve the reliability of block propagation, we introduce a reputation mechanism based on the subjective logic model, including the local and recommended opinions to calculate the miner reputation value. Moreover, considering that the GAT possesses the excellent ability to process graph-structured data, we utilize the GAT with reinforcement learning to obtain the optimal block propagation trajectory. Numerical results demonstrate that the proposed scheme exhibits the most outstanding block propagation efficiency and reliability compared with traditional routing mechanisms.
\end{abstract}


\section{Introduction}
Driven by the increasing interest in cryptocurrencies and the advancement of blockchain technologies, Web 3.0 is emerging to enable users to securely manage data without a centralized authority\cite{chen2022digital}. Since the advent of the World Wide Web, there have been three generations of the Web. Specifically, Web 1.0 was featured by static web pages and limited user engagement\cite{chen2022digital}. Web 2.0 is a paradigm shift in how the internet is used, which is characterized by interactivity and social connectivity\cite{chen2022digital}. Unlike Web 2.0 and Web 1.0, Web 3.0 is envisioned as the blockchain-based Internet, which holds the potential to revolutionize the modern Internet in two critical aspects. Firstly, Web 3.0 transcends being merely a readable and writable network, which empowers users to assert dominion over their data and assets\cite{xu2023quantum}. Secondly, Web 3.0 embodies a nascent economic system collaboratively constructed and shared by both end-users and creators\cite{xu2023quantum}.

 With the incredible ability to empower secure, transparent, and tamper-proof data transactions through decentralized ledgers, blockchains have attracted significant attention from both academia and industry\cite{wen2022optimal}. Based on consensus mechanisms and encryption technologies, blockchains possess the ability to facilitate the creation of cryptocurrencies and drive the development of decentralized applications and smart contracts, assuming a pivotal role across various domains, such as decentralized finance\cite{werner2022sok}, Metaverse\cite{kang2023blockchain}, and Internet of Vehicles\cite{shen2022secure}. In Web 3.0, a significant volume of transactions involving digital products, such as Non-Fungible Tokens, transpires among users,  which poses challenges to data security. To this end, blockchain-enabled Web 3.0 emerges as a necessary strategy to ensure secure storage and efficient management of these transactions\cite{xu2023quantum}. Furthermore, the blockchain-enabled Web 3.0 facilitates decentralized data exchange and sharing, which enables participants to circumvent the dominance of tech giants and gain control over user-generated content\cite{lin2023blockchain}. Therefore, blockchains are widely regarded as fundamental and indispensable technologies for the progress of Web 3.0.

Although Web 3.0 is more accessible and intelligent than previous generations, it still faces some challenges for future popularization and development. One of the main challenges is improving blockchain performance\cite{xu2023quantum}. For one thing, in public blockchains, a new block is randomly broadcast in the miner network for validation, which may lead to large overall propagation time\cite{jiang2021taming}. When block propagation time in the miner network becomes excessively prolonged, it can lead to an excessive number of forks and insufficient signature collections, ultimately resulting in the failure of transaction verification\cite{jiang2021taming, wen2022optimal}. Additionally, prolonged propagation delay may significantly extend the block generation interval. For another thing, during the block propagation process, malicious miners can easily access blockchains and issue attacks, such as Sybil attacks, double-spending attacks, and selfish mining, which results in financial losses and privacy breaches, substantially eroding blockchain reliability. Therefore, to effectively improve blockchain performance, reliably optimizing block propagation is critical\cite{wen2022optimal}. Some efforts have been conducted to optimize block propagation\cite{wen2022optimal, lin2023blockchain, xu2023quantum, jiang2021taming}. However, they do not consider \textit{how to measure the block propagation efficiency in public blockchains and how to quantify the reliability of miners in the block propagation process for enabling reliable block propagation optimization.}

To address the above challenges, in this paper, we design a Graph Attention Network (GAT)-based reliable block propagation optimization framework for blockchain-enabled Web 3.0. Specifically, since Age of Information (AoI) as a well-established metric can capture information freshness\cite{wen2023freshness}, we apply the Age of Block (AoB), which is similar to AoI, to measure block propagation efficiency in public blockchains. Then, to achieve reliable block propagation optimization, we propose a miner reputation mechanism based on the subjective logic model\cite{zhong2023blockchain}, including the local and recommended opinions to measure miner reliability in public blockchains. Furthermore, we utilize the GAT with Reinforcement Learning (RL) to extract the relational presentation between each miner and its adjacent miner in a graph-structured miner network, thereby facilitating the formulation of the miner selection process. By integrating the aforementioned methods, the framework can ultimately derive the optimal block propagation trajectory.
The contributions of this paper are summarized as follows:
\begin{itemize}
    \item We adopt the AoB as a data-freshness metric to measure block propagation efficiency in public blockchains, which has a specific consideration of the block consensus mechanism. For example, the proposed AoB innovatively provides a comprehensive examination of the procedures of getdata message arrival time interval, block waiting time, and block propagation time. Moreover, unlike \cite{wen2023task}, we infer the specific formulas considering the characteristics of block propagation in public blockchains.
    \item To ensure the reliability of block propagation, we propose a miner reputation mechanism based on the subjective logic model, where the subjective logic model as a widely adopted mathematical tool utilizes the local and recommended miner opinions to effectively calculate miner reputation values in the miner network. Moreover, the miner reputation mechanism has been specifically designed in both local and recommended opinions to improve the reputation calculation.
    \item To achieve reliable block propagation optimization, we propose a GAT-based block propagation optimization model to delve into more intricate aspects during the block consensus process than existing consensus mechanisms, in which we design a novel GAT-based architecture with RL to minimize the overall AoB of block propagation given the reputation constraint, thereby obtaining the optimal block propagation trajectory. Moreover, considering the consensus mechanism of public blockchains (i.e., Proof of Work), the optimization conducted by the GAT model is based on a specific number of miners.
    \item Compared with Greedy and Gossip mechanisms, our model exhibits outstanding performance and achieves the most efficient and reliable block propagation. For example, when the number of miners is set to $49$, the overall AoB of our model is $2.4\%-22.6\%$ lower than that of Greedy and Gossip mechanisms. Moreover, the total value of miner reputations of our model is $7.4\%-34.77\%$ higher than that of Greedy and Gossip mechanisms.
\end{itemize}

 \begin{figure*}[t]
 \centering     
 \includegraphics[width=0.95\textwidth]{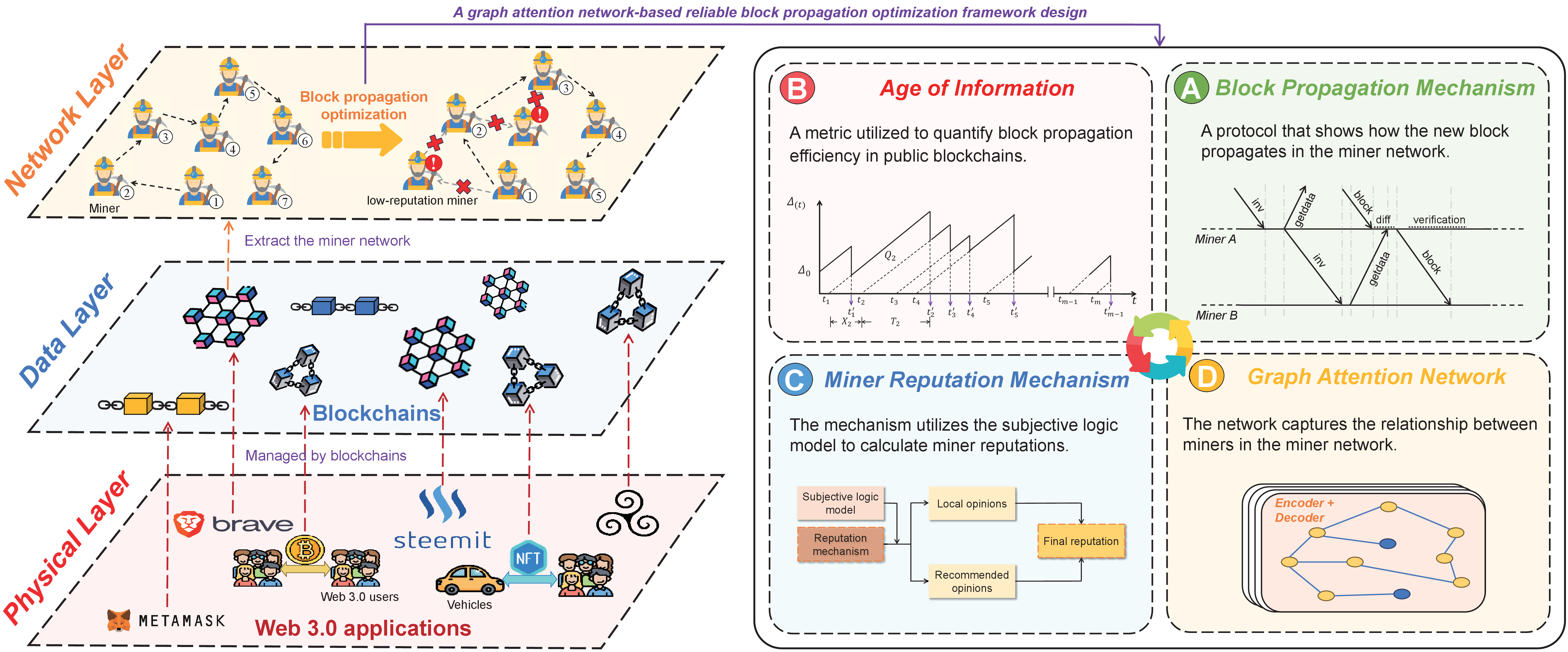}     
 \caption{A GAT-based reliable block propagation optimization framework for blockchain-enabled Web 3.0.}\label{framework}     
\end{figure*}

The remainder of the paper is organized as follows: Section \ref{II} reviews related literature. In Section \ref{III}, we propose the system model including the block propagation mechanism and the proposed framework. In Section \ref{IV}, we formulate the problem of minimizing the overall AoB of block propagation with the given reputation constraint. In Section \ref{V}, we introduce the GAT with the encoder-decoder architecture. In Section \ref{VI}, we introduce RL with a rollout baseline to train the network. Section \ref{VII} presents the numerical results of the proposed schemes. Finally, the paper concludes with Section \ref{VIII}.
 
\section{Related Work} \label{II}
\renewcommand{\arraystretch}{1.4}
\begin{table}[t]
\caption{Mathematical Notations}
\begin{tabular}{m{0.8cm}|m{6.8cm}} 
\toprule[1pt]
\hline
\multicolumn{1}{c|}{\textbf{Notation}}  & \multicolumn{1}{c}{\textbf{Definition}} \\ \hline
$\Gamma_i$ &  Block propagation time between miner $(i-1)$ and miner $i$ \\ \hline
$X_i$ & Time elapsed from miner $(i-1)$ sending a getdata message to miner $i$ sending a getdata message \\ \hline
$D_i$ & Time elapsed from miner $i$ sending a getdata message at time $t_i$ to successfully receive the block at $t_i^{'}$  \\ \hline
$W_i$ & Block waiting time, as a part of the system time $D_i$  \\ \hline
$\Delta$ & Average AoB over the time range (0, $\delta$)   \\ \hline
$\alpha_{i\to j}^{t_k}$ & Number of positive interactions between miners\\ \hline
$\beta_{i\to j}^{t_k}$& Number of negative interactions between miners  \\ \hline

$T_{i\to j}^{t_k}$ & Trust of the local reputation of miner
$i$ to miner $j$ in the time slot $t_k$\\ \hline
$F_{i\to j}^{t_k}$ & Distrust of the local reputation of miner
$i$ to miner $j$ in the time slot $t_k$ \\ \hline
$U_{i\to j}^{t_k}$ & Uncertainty of the local reputation of miner
$i$ to miner $j$ in the time slot $t_k$ \\ \hline
$R_{i\to j}^{final}$ & Final reputation value of miner
$i$ to miner $j$\\ \hline
$\boldsymbol\theta$ & Network parameters of the proposed GAT model\\ \hline
$\boldsymbol q_{iy}^{(r)}$ & Query embedding of miner $i$ in the $y$-th head of the $r$-th GAT layer\\\hline
$\boldsymbol k_{iy}^{(r)}$ & Key embedding of miner $i$ in the $y$-th head of the $r$-th GAT layer\\\hline
$\boldsymbol v_{iy}^{(r)}$ & Value embedding of miner $i$ in the $y$-th head of the $r$-th GAT layer\\
\hline
$\boldsymbol u_{ijy}^{(r)}$ & Compatibility embedding from miner $i$ to miner $j$ in the multi-head attention layer of the GAT model\\\hline

\bottomrule[1pt]
\end{tabular}
\end{table}

\subsection{Blockchain-enabled Web 3.0}
Web 3.0, also regarded as the semantic web, represents the cutting edge of web development, driven by Artificial Intelligence (AI), Internet of Things, and blockchain technologies. Nowadays, the intertwining of blockchain technologies with Web 3.0 forms a symbiotic relationship, giving rise to blockchain-enabled Web 3.0, which is effective for the transparent, traceable, and decentralized recording of on-chain content. Therefore, research aimed at enhancing blockchain performance in  Web 3.0 holds unique significance, serving as a key motivation for our work. Some efforts have been conducted to achieve blockchain-enabled Web 3.0\cite{xu2023quantum, lin2023blockchain, zhang2023ai}. For example, the authors in\cite{xu2023quantum} introduced a Web 3.0 framework powered by quantum blockchain technologies and discussed potential challenges and applications of integrating quantum blockchain in Web 3.0\cite{xu2023quantum}. The authors in \cite{lin2023blockchain} proposed a semantic exchange framework leveraging blockchain technologies and discussed the relationship between semantic communication and blockchain for Web 3.0\cite{lin2023blockchain}. The authors in\cite{zhang2023ai} developed an innovative architecture for Web 3.0, integrating AI and blockchain to establish a metaverse-based framework that aims to tackle the existing privacy and security challenges of the current Web 2.0 while improving user experience and data control. Moreover, the authors emphasized the importance of blockchain technologies in the construction of Web 3.0\cite{zhang2023ai}. However, the majority of existing works do not delve into the optimization of blockchain performance as a means to enhance Web 3.0 performance. Therefore, it is important to optimize blockchain performance for Web 3.0, especially optimizing block propagation.  

\subsection{Graph Attention Network} 
In recent years, GAT as a subtype of graph neural networks, has garnered considerable attention, including molecules, social networks, product recommendations, computer programs, and more\cite{velivckovic2017graph}. GAT is a class of deep learning models specially designed for processing graph-structured data, known for their exceptional adaptability and efficiency across various applications\cite{lei2022solve}. For example, the authors in \cite{lei2022solve} proposed a novel residual edge-graph attention network that combines edge and node information to address combinatorial optimization problems\cite{lei2022solve}. The authors in\cite{zhang2023spatio} employed a foundational three-layer GAT network to capture spatial structural information based on the dynamic aggregation in the road network. Furthermore, when compared with the graph convolutional network, the GAT model significantly enhances the reliability of traffic flow prediction\cite{zhang2023spatio}. The authors in\cite{huang2022gat} introduced a performance prediction model for multipath routing, utilizing the GAT, which models the function and structure of data to predict the network stability and throughput\cite{huang2022gat}. However, the majority of the existing works do not take into account the utilization of GAT to capture the miner network information, such as the relationship between adjacent miners, thereby optimizing block propagation. 

\subsection{Block Propagation Optimization}
Block propagation optimization has been a prominent research topic for improving blockchain performance. There are numerous researchers investigating this issue, and their efforts can be broadly categorized into three main directions: 1) Optimizing block verification: the authors in\cite{Rwang2023block} proposed a block verification mechanism based on zero-knowledge proof and smart contracts, which improves both the speed of data block verification and privacy protection\cite{Rwang2023block}. The authors in\cite{rzhai2023tvs} introduced a secure verification scheme for real-time office documents leveraging blockchain technologies, thereby enhancing the efficiency and security in block verification\cite{rzhai2023tvs}.  2) Optimizing blockchain network topology: the authors in\cite{rhaselbarth2023blockchain} proposed a two-layer blockchain topology to overcome transaction time issues and scalability for market procurement of reactive power\cite{rhaselbarth2023blockchain}. The authors in\cite{rdeshpande2022efficient} employed the software-defined networking paradigm to optimize the control of the peer-to-peer network topology in public blockchains, resulting in a significant reduction in node resource consumption\cite{rdeshpande2022efficient}. 3) Optimizing the propagation behavior of miners:  the authors in\cite{rchen2023prevention} introduced a method for preventing block withholding attacks, which utilizes a credit model, a punishment mechanism, and the behavior reward to evaluate the contribution of miners\cite{rchen2023prevention}. However, few works have been conducted on combining block propagation efficiency and miner reliability based on the GAT when optimizing block propagation. In Web 3.0, efficient and reliable blockchain technology can significantly promote further development in this ecosystem, giving users a better experience. Therefore, it is still challenging to leverage the GAT to achieve efficient and reliable block propagation optimization. 
 
Motivated by the aforementioned research gaps, we propose a GAT-based reliable block propagation optimization framework for the performance enhancement of blockchain-enabled Web 3.0, in which we apply the AoB to quantify block propagation efficiency and propose a miner reputation mechanism to ensure the reliability of block propagation.

\section{GAT-based Reliable Block Propagation Optimization Framework for Web 3.0} \label{III}

In this section, we first briefly introduce the block propagation mechanism in public blockchains\cite{Adecker2013information}. Then, we provide a comprehensive description of the proposed GAT-based reliable block propagation optimization framework for blockchain-enabled Web 3.0, as illustrated in Fig. \ref{framework}. 

\subsection{Block Propagation Mechanism}
For the block propagation in public blockchains, we introduce the block propagation mechanism which is similar to the traditional gossip protocol\cite{Adecker2013information, chen2023accelerated}. However, for the optimization of specific block propagation trajectories, the introduced block propagation mechanism directs the propagation of a new block to a single miner, departing from the flooding propagation characteristic of the gossip. The detailed process is depicted in Part A of Fig. \ref{framework}. To prevent redundant transmissions to miners that have already received the new block, the new block is not forwarded directly\cite{Adecker2013information}. Instead, a miner would send an \textit{inv} message containing the hash value of the new block to its adjacent miner before propagating the block. If the adjacent miner has never received the block, the adjacent miner will issue a \textit{getdata} message to request the new block \cite{Adecker2013information, Bzhang2022speedingBLCOKPROPAGATION}, and the getdata message is then queued until the block has been received and the difficulty check has been done\cite{Adecker2013information}. To speed up block propagation, the adjacent miner sending the getdata message also sends an inv message to its adjacent miner. Additionally, the block verification process is divided into two phases, i.e., the block difficulty check and transaction validation \cite{Adecker2013information}. Once the block difficulty check is completed, the block is sent out without further transaction validation, aiming to minimize processing time \cite{Adecker2013information}. 

\subsection{Framework Design}
As shown in Fig. \ref{framework}, the proposed GAT-based reliable block propagation optimization framework for blockchain-enabled Web 3.0 is structured into three layers, i.e., a physical layer, a data layer, and a network layer. Specifically, the first layer, functioning as the foundation of the proposed framework, is the physical layer, which exists in various applications in Web 3.0, such as decentralized autonomous organizations and decentralized marketplaces for digital products\cite{chen2022digital}. The second layer is the data layer, where blockchains provide secure and reliable data management for Web 3.0 applications\cite{xu2023quantum}. The final layer is the network layer, which is exacted from the blockchains of the data layer. Especially, the network layer introduces the proposed framework design for achieving block propagation optimization, including four parts. Part A presents a block propagation protocol that shows how a new block propagates between miners in the miner network\cite{Adecker2013information}. Part B introduces the AoB utilized to measure the block propagation efficiency in public blockchains. Part C focuses on the miner reputation mechanism, which utilizes the subjective logic model to calculate miner reputations for ensuring reliable block propagation\cite{zhong2023blockchain}. Part D shows the GAT, which works directly on graph-structured data and possesses the excellent ability to capture the relationships between miners in the miner network\cite{Gkool2018GNNyuanxing}. The proposed framework aims to achieve efficient and reliable block propagation optimization, thereby enhancing the overall performance of Web 3.0 systems.

\section{Problem Formulation for Efficient and Reliable Block Propagation}\label{IV}
In this section, we first introduce the AoB to quantify block propagation efficiency in public blockchains. Then, we propose a miner reputation mechanism based on the subjective logic model. In this paper, we consider that there is a set $\mathcal{M} = \left\{1,\ldots, i,\ldots,j,\ldots, M\right\}$ of $M$ miners with the nature of random mobility in the miner network, and the block propagation process can be considered an M/M/1 system following a first-come-first-served policy.

\subsection{Age of Block for Block Propagation}

Since AoI as an effective data-freshness metric, has been widely utilized in remote system status, which is denoted as the elapsed time from the generation of the latest received status update\cite{Akosta2017age}. In this paper, we utilize the AoB to quantify block propagation efficiency motivated by \cite{Akosta2017age, wen2023task}, which is denoted as the time elapsed from a getdata message sent by a miner to its successful receipt of the new block, consisting of procedures of getdata message arrival time interval, block waiting time, and block propagation time\cite{wen2023task}. We denote $t_i$ as the time that miner $i$ sends a getdata message and $t_i^{'}$ as the time that miner $i$ successfully receives the block. Thus, the instantaneous AoB of block propagation between adjacent miners at any time $t$ is expressed as \cite{Akosta2017age}
\begin{equation}\label{delta}
    \Delta(t)=t-t_{I(t)}, 
\end{equation}
where $I(t)=\max\{i|t_{i}^{'}\leq t,i\in \mathcal{M}\}$ represents the index of the miner that recently received the block. Based on (\ref{delta}), the overall AoB over the time range $(0,\delta)$ is given by 
\begin{equation} \label{Delta_T}
    \Delta_{\delta}=\int_0^{\delta}\Delta(t)\mathrm{d}t,
\end{equation}
and the average AoB over the time range $(0,{\delta})$ is expressed as 
\begin{equation} \label{2}
    {\Delta}=\frac{1}{\delta}\int_0^{\delta}\Delta(t)\mathrm{d}t.
\end{equation}

Now, we show the characterization of the average AoB of block propagation between adjacent miners in public blockchains. We define the $i$-th time interval, denoted by $X_i$, as the time elapsed from miner $(i-1)$ sending the getdata message to miner $i$ sending the getdata message, which is given by
\begin{equation} \label{X_i}
    X_i=t_i-t_{i-1}.
\end{equation}
We consider $X_i$ as an exponentially distributed random variable\cite{Akosta2017age}, i.e., $X_i\sim Exp(\mu)$, where $\mathbb{E}[X_i]=1/\mu$ and $\mu$ represents the average getdata delivery ratio of miners. Then, we denote the system time of miner $i$ as $D_i$, equaling the elapsed time from miner $i$ sending a getdata message at time $t_i$ to successfully receive the block at $t_i^{'}$, which is given by
\begin{equation} \label{D_i}
    D_i=t_i^{'}-t_{i},
\end{equation}
where system time $D_i$ consists of block waiting time $W_i$ of miner $i$ and block propagation time $P_i$ from miner $(i-1)$ to miner $i$. As shown in Fig. \ref{system}, $Q_i$ signifies the geometric area of a trapezoidal configuration, arising from the difference between two isosceles right-angled triangles. Based on (\ref{X_i}) and (\ref{D_i}), $Q_i$ can be expressed as\cite{Akosta2017age}
\begin{equation} \label{Q_i}
    Q_i=0.5{(D_i+X_i)}^2-0.5{D_i}^2=X_i D_i+0.5{X_i}^2.
\end{equation}
Therefore, the AoB of block propagation between miner $(i-1)$ and miner $i$ in public blockchains is given by\cite{Akosta2017age}
\begin{equation}\label{AoB}
    \Delta_i= \frac{\mathbb{E}[Q_i]}{\mathbb{E}[X_i]} = \frac{\mathbb{E}[X_iD_i]+0.5\mathbb{E}[X_i^2]}{\mathbb{E}[X_i]}.
\end{equation}

Considering the system time $D_i$ of miner $i$, we divide the system time $D_i$ into block waiting time $W_i$ and block propagation time $P_i$, which is denoted as
\begin{equation}
    D_i=W_i+P_i.
\end{equation}
Since $P_i$ is independent of time interval $X_i$, we can obtain
\begin{equation}
   \mathbb E[X_iD_i]=\mathbb E[X_i(W_i+P_i)],
\end{equation}
\begin{equation} \label{E[T][Y]}
	\mathbb{E}[D_iX_i]=\mathbb{E}[W_iX_i]+\mathbb{E}[P_i]\mathbb{E}[X_i].
\end{equation}
According to \cite{Akosta2017age}, we can express the block waiting time of miner $i$ as $W_i=(D_{i-1}-X_i)^+$. Given specific interarrival time $X_i = x$, the expected block waiting time is \cite{Akosta2017age}
\begin{equation} \label{WW}
	\mathbb{E}[W_i|X_i=x]=\mathbb{E}[(D-x)^+]=\int_x^\infty(t-x)f_D(t)\mathrm{d}t,
\end{equation}
where $f_D(t)$ is the Probability Density Function (PDF) of system time $D$. It is shown that the exponential distribution is a reasonable model for block propagation time\cite{rovira2019optimizing}, i.e., $P_i \sim Exp(1/\Gamma_i)$, where $\Gamma_i$ is the block propagation time between miner $(i-1)$ and miner $i$, which is calculated using the Shannon formula as follows \cite{zhong2023blockchain}:
\begin{equation} \label{t_p}
    \Gamma_i=\frac{B_{block}}{b\log_2\Big(1+\frac{\rho_sc^0d_{i-1,i}^{-\varepsilon}}{N_0b}\Big)},
\end{equation}
where $B_{block}$, $b$, $\rho_s$, ${d_{i-1, i}}$, $c^0$, $\varepsilon$, and $N_0$ represent the block size, the channel bandwidth between adjacent miners, the transmit power of miners, the distance between miner $(i-1)$ and miner $i$, the unit channel power gain, the path-loss coefficient, and the noise power density, respectively. For the system time $D_i$ of miner $i$, $f_{D_i}(t)$ is given by\cite{Akosta2017age}
\begin{equation}\label{ffT}
	f_{D_i}(t)=\bigg(\frac{1}{\Gamma_i}-\mu\bigg)e^{(\mu-\frac{1}{\Gamma_i})t},\: t\geq 0.
\end{equation}
Furthermore, using iterated expectation and the exponential $(\mu)$ PDF of $X_i$\cite{Akosta2017age}, $\mathbb{E}[W_iX_i]$ can be denoted as
\begin{equation}
\begin{aligned}
    \mathbb{E}[W_iX_i]&=\int_0^\infty x\mathbb{E}[W_i|X_i=x]f_{X_i}(x)\mathrm{d}x\\
    &=\frac\mu{({\frac{1}{\Gamma_i}})^3-({\frac{1}{\Gamma_i}})^2\mu}. 
\end{aligned}
\end{equation}
Therefore, based on (\ref{AoB}), the AoB between miner $(i-1)$ and miner $i$ can be given by
\begin{equation}\label{Delta_final}
\begin{aligned}\Delta_i={\Gamma_i}+\frac1\mu+\frac{\mu \Gamma_i^3}{1-\mu \Gamma_i}.\end{aligned}
\end{equation}
(\ref{Delta_final}) demonstrates that low overall AoB means the comprehensive reduction of the time of block propagation process, including the getdata message arrival time interval, the block waiting time, and the block propagation time, which indicates the enhanced block propagation
 efficiency. Furthermore, the PDF of a block to be mined can be given by\cite{shahsavari2019theoretical}
\begin{equation}
f(t;\mu)=\mu e^{-\mu t}.\label{ft}
\end{equation}
Based on (\ref{ft}), we can obtain the block fork probability, which is a function of block propagation time and the average getdata delivery ratio of miners, given by \cite{shahsavari2019theoretical}
\begin{equation}F(t)=P(X_i\leq {\Gamma})=\int_{0}^{{\Gamma}}f(t)\mathrm{d}t=1-e^{-\mu{{\Gamma}}},\label{fyyy}\end{equation}where $\Gamma$ is the overall block propagation time. Thus, combining (\ref{fyyy}) and (\ref{fu}), it can be inferred that a low overall AoB can reduce the block fork probability when $\mu$ is a fixed value.
 
\subsection{Reliable Block Propagation based on the Subjective Logic Model} 
The subjective logic model, which utilizes logic and recommended opinions to express subjective beliefs,  is a comprehensive model of subjective beliefs evaluation. Moreover, the miner interactions in the blockchain can be considered as a form of group behavior, which can be calculated by direct and indirect opinions\cite{zhong2023blockchain}. Therefore, for a better fusion of probabilistic information, we utilize the subjective logic model as an effective tool to quantify the reliability of miners in the blockchain network.

 Based on the subjective logic model, the subjective logic divides miner reputations into three levels, i.e., trust, distrust, and uncertainty\cite{jiang2022reliable, zhong2023blockchain}. For the miner $j$, its compositive reputation opinions consist of the indirect reputation opinions of other adjacent miners and the direct reputation opinions of miner $i$ that directly interact with the miner $j$. Especially, the direct reputation opinions of miner $i$ are considered to be local reputation opinions, and the indirect reputation opinions of other adjacent miners are considered to be recommended reputation opinions\cite{jiang2022reliable}.


\subsubsection{Local opinions for subjective logic}
We divide the effective interaction period with multiple block consensuses into a series of time windows as $\{t_1,\ldots,t_k,\ldots,t_K\}$. In the subjective logic model, the local reputation opinions of miner $i$ to miner $j$ in the time slot $t_k$ are represented as a tuple vector $\Phi_{i\to j}^{t_k}:=\{T_{i\to j}^{t_k}, F_{i\to j}^{t_k}, U_{i\to j}^{t_k}\}$\cite{zhong2023blockchain}, where $T_{i\to j}^{t_k}$, $F_{i\to j}^{t_k}$, and $U_{i\to j}^{t_k}$ represent trust, distrust, and uncertainty, respectively. Here $T_{i\to j}^{t_k}$, $F_{i\to j}^{t_k}$, and $U_{i\to j}^{t_k}\in[0,1]$ and $T_{i\rightarrow j}^{t_k} + F_{i\rightarrow j}^{t_k} + U_{i\rightarrow j}^{t_k}=1$. Without loss of generality, we consider that the positive interaction of miner $i$ with miner $j$ includes miner $j$ timely sending getdata messages to miner $i$ or being willing to forward the new block to its suitable adjacent miners, and the interactions between miner $i$ and miner $j$ result in the positive mapping of trust and the negative mapping of distrust. Therefore, we can obtain $\Phi_{i\to j}^{t_k}$ as
 \begin{equation} \label{initial reputation}
      \begin{cases}T_{i\to j}^{t_k}=\left(1-U_{i\to j}^{t_k}\right)\frac{\alpha_{i\to j}^{t_k}}{(\alpha_{i\to j}^{t_k}+\beta_{i\to j}^{t_k})},\\F_{i\to j}^{t_k}=\left(1-U_{i\to j}^{t_k}\right)\frac{\beta_{i\to j}^{t_k}}{(\alpha_{i\to j}^{t_k}+\beta_{i\to j}^{t_k})},\\U_{i\to j}^{t_k}=1-q_{i\to j}^{t_k},\end{cases}
 \end{equation}
where $\alpha_{i\to j}^{t_k}$, $\beta_{i\to j}^{t_k}$, and $q_{i\to j}^{t_k}$ denote the number of positive miner interactions, the number of negative miner interactions, and the probability of successful block transmission, in the time window $t_k$, respectively. 
Note that a large value of  $q_{i\to j}^{t_k}$ indicates the good quality of the communication link between miner $i$ and miner $j$ during block propagation. Therefore, the uncertainty of block propagation $U_{i\to j}^{t_k}$ is associated with $q_{i\to j}^{t_k}$. Specifically, if there exist certain factors that prevent the block from propagating properly between adjacent miners during the block propagation process, the failure probability of block transmission will increase, leading to the higher $U_{i\to j}^{t_k}$. Based on the above analysis, the local reputation value of miner $i$ to miner $j$ in the time slot $t_k$ is expressed as\cite{jiang2022reliable, zhong2023blockchain}
\begin{equation}\label{R_tk} 
    R_{i\to j}^{t_k}=T_{i\to j}^{t_k}+\eta U_{i\to j}^{t_k},
\end{equation}
where $\eta\in [0,1]$ represents the degree of uncertainty effect on miner reputations\cite{jiang2022reliable}. To improve the reliability of block propagation, we further explore the miner reputation during block propagation by considering two aspects, i.e., the impact of positive and negative interactions and the freshness of interactions between adjacent miners, which can more accurately quantify the reliability of miners\cite{zhong2023blockchain}.
\begin{itemize}
    \item \textbf{\emph{Interaction effects:}} 
    To curb the negative interactions between adjacent miners during block propagation, greater weight should be assigned to $\beta_{i\to j}^{t_k}$. We introduce $\xi\in(1,\infty)$ as the weight of negative interactions. A larger value of $\xi$ indicates that the influence of a negative interaction on the miner's reputation is more significant compared to that of a positive interaction. Therefore, (\ref{initial reputation}) are modified to
\begin{equation}
    \begin{cases}T_{i\to j}^{t_k}=\left(1-U_{i\to j}^{t_k}\right)\frac{\alpha_{i\to j}^{t_k}}{(\alpha_{i\to j}^{t_k}+\xi\beta_{i\to j}^{t_k})},\\F_{i\to j}^{t_k}=\left(1-U_{i\to j}^{t_k}\right)\frac{\xi\beta_{i\to j}^{t_k}}{(\alpha_{i\to j}^{t_k}+\xi\beta_{i\to j}^{t_k})},\\U_{i\to j}^{t_k}=1-q_{i\to j}^{t_k}.\end{cases}
\end{equation}

    \item \textbf{\emph{Interaction freshness:}} Since recent interactions, characterized by high freshness, have a greater impact on the miner reputation compared to past interactions, to reflect the impact of interaction freshness on the miner reputation, we introduce an \textit{interaction freshness function} as $f(t)=\lambda\cdot\ln t$, where $\lambda\in(0,1)$ represents a temporal enhancement factor reflecting the influence weight of interaction freshness and the local reputation opinion of miner $i$ to miner $j$ can be obtained as
\begin{equation}
    \begin{cases}T_{i\to j}^{local}=\frac{\sum_{k=1}^{K}f(t_k)T_{i\to j}^{t_k}}{\sum_{k=1}^{K}f(t_k)},\\F_{i\to j}^{local}=\frac{\sum_{k=1}^{K}f(t_k)F_{i\to j}^{t_k}}{\sum_{k=1}^{K}f(t_k)},\\U_{i\to j}^{local}=\frac{\sum_{k=1}^{K}f(t_k)U_{i\to j}^{t_k}}{\sum_{k=1}^{K}f(t_k)}.\end{cases}
\end{equation}
Therefore, the local reputation of miner $i$ to miner $j$ $R_{i\to j}^{local}$, representing the expected belief of miner $i$ that miner $j$ is reliable and willing to forward the block to its suitable adjacent miners during the block propagation process, is given by\cite{Rkang2019toward}
\begin{equation}
    R_{i\to j}^{local}=T_{i\to j}^{local}+\eta U_{i\to j}^{local}.
\end{equation}
\end{itemize}

\subsubsection{Recommended opinions for subjective logic}
In the miner network, each miner has its own set of interacting peers. It is obvious that frequent interactions between adjacent miners can enhance the degree of their recommended opinions. The reason is that frequent interactions indicate a high level of mutual recognition and great pre-existing knowledge of each other between adjacent miners. We consider a set $\mathcal{S}=\{1,\ldots,s,\ldots, S\}\subset\mathcal{M}$ of $S$ recommenders and apply an interaction frequency metric within miners and their recommenders\cite{zhong2023blockchain}. 

Especially, considering that miners who engage in more positive interactions should obtain a larger proportion of recommendations, we introduce $\gamma^{rec}\in(1, \infty)$ as an enhancement factor reflecting the influence weight of positive interactions on recommendations. Moreover, for recommender $s$ and miner $j$, the interaction frequency is defined as the ratio of the interactions between recommender $s$ and miner $j$ to the average value of interactions between recommender $s$ and miners, which is denoted as $IF_{s\to j}=H_{s\to j}/\overline{H}_{s}$, where $H_{s\to j}=\gamma^{rec}\alpha_{s\to j}+\beta_{s\to j}$ and $\overline{H}_s=\left(\frac{1}{M}\right)\sum_{j=1}^{M}H_{s\to j}$ \cite{Rkang2019toward}. Furthermore, the weight of recommended opinions of recommender $s$ is defined as $\omega_{s\to j}=\delta_{s\to j}\cdot IF_{s\to j}$, where $\delta_{s\to j}$ is a pre-defined coefficient for the calculation of the recommended reputation. Therefore, the recommended reputation opinion of recommender $s$ to miner $j$ can be obtained as 
\begin{equation}
    \begin{cases}T_{s\to j}^{rec}=\frac{\sum_{s\in \mathcal{S}}\omega_{s\to j}T_{s\to j}^{local}}{\sum_{s\in \mathcal{S}}\omega_{s\to j}},\\F_{s\to j}^{rec}=\frac{\sum_{s\in \mathcal{S}}\omega_{s\to j}F_{s\to j}^{local}}{\sum_{s\in \mathcal{S}}\omega_{s\to j}},\\U_{s\to j}^{rec}=\frac{\sum_{s\in \mathcal{S}}\omega_{s\to j}U_{s\to j}^{local}}{\sum_{s\in \mathcal{S}}\omega_{s\to j}}.\end{cases}
\end{equation}
\subsubsection{Final reputation for subjective logic}
For calculating final miner reputations, it is important to consider the weights of local reputation opinions reasonably and recommended reputation opinions, thereby ensuring the final reputation of miners is more accurate and trustworthy. Based on the above analysis, the final reputation opinion of miner $i$ to miner $j$ is composed of a tuple vector $\Phi_{i\to j}^{final}:=\{T_{i\to j}^{final}, F_{i\to j}^{final}, U_{i\to j}^{final}\}$, which is given by
\begin{equation}
    \begin{cases}T_{i\to j}^{final}=\frac{T_{i\to j}^{local}U_{s\to j}^{rec}+T_{s\to j}^{rec}U_{i\to j}^{local}}{U_{i\to j}^{local}+U_{s\to j}^{rec}-U_{s\to j}^{rec}U_{i\to j}^{local}},\\F_{i\to j}^{final}=\frac{F_{i\to j}^{local}U_{s\to j}^{rec}+F_{s\to j}^{rec}U_{i\to j}^{local}}{U_{i\to j}^{local}+U_{s\to j}^{rec}-U_{s\to j}^{rec}U_{i\to j}^{local}},\\U_{i\to j}^{final}=\frac{U_{s\to j}^{rec}U_{i\to j}^{local}}{U_{i\to j}^{local}+U_{s\to j}^{rec}-U_{s\to j}^{rec}U_{i\to j}^{local}}.&\end{cases}
\end{equation}
Based on (\ref{R_tk}), the final reputation value of miner $i$ to miner $j$ is expressed as
\begin{equation}
    R_{i\to j}^{final}=T_{i\to j}^{final}+\eta U_{i\to j}^{final}.
\end{equation}
\section{Graph Attention Network-based Block Propagation Model} \label{V}
 \begin{figure*}[t]\centering     \includegraphics[width=1\textwidth]{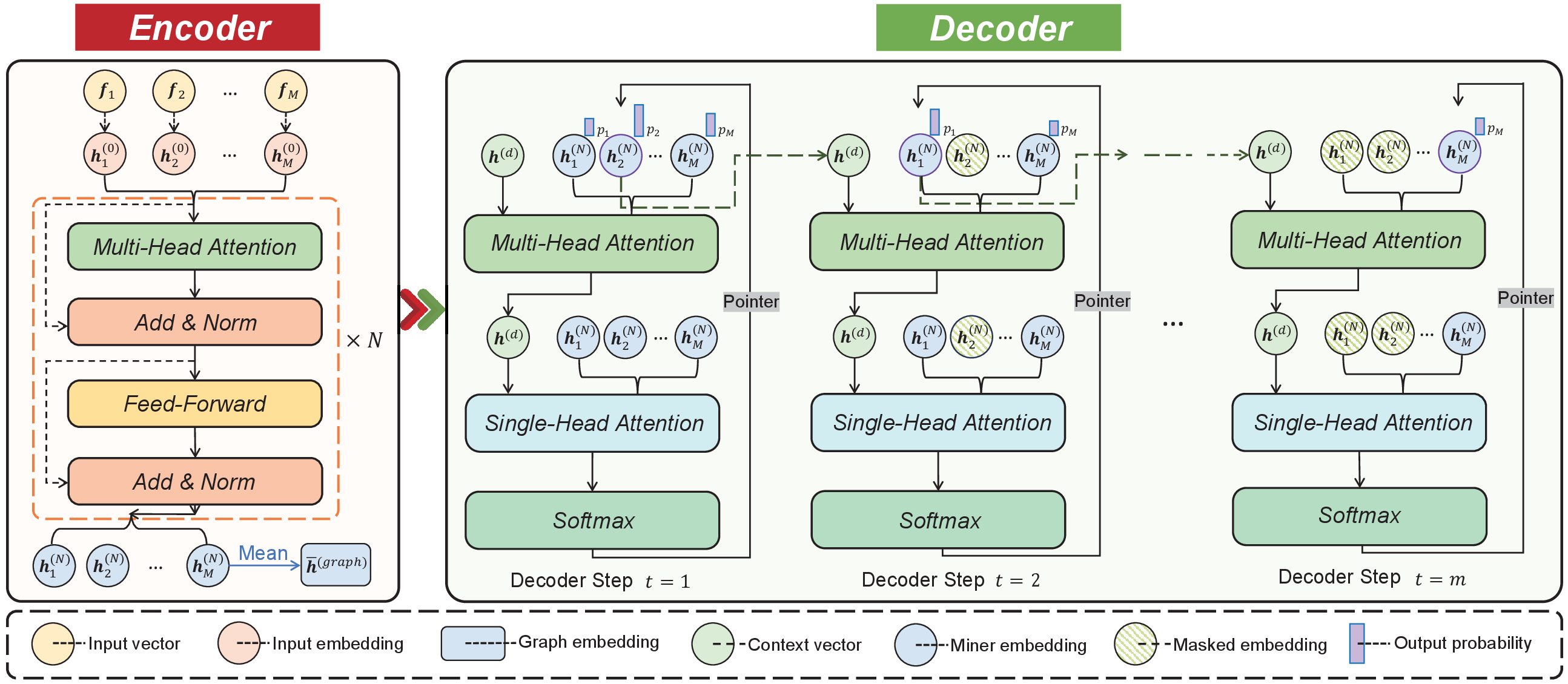}     \caption{The encoder and decoder architectures in the GAT-based block propagation optimization model.}  \label{system}     
\end{figure*}
\subsection{Problem Formulation}
In this paper, we focus on achieving reliable block propagation optimization in public blockchains by obtaining the solution path of optimal block propagation $\boldsymbol{\pi}=\left(\pi_{1},\ldots,\pi_{m}\right)$, where $\pi_{t}\in\{1,\ldots,m\}\subseteq \mathcal{M}$ and $\pi_t\neq\pi_{t^{\prime}},\forall t\neq t^{\prime}$. Note that $\boldsymbol{\pi}$ is the miner set of the optimal block propagation trajectory and $\pi_1$ represents the miner that obtains the bookkeeping right of this block consensus. 

Based on the consensus mechanism of public blockchains (i.e., Proof of Work), we define $m=\lfloor{M\times r^{ratio}}\rfloor$ as the number of miners that participate in block validation, where $\lfloor \cdot\rfloor$ denotes the floor function and $r^{ratio}$ represents the proportion of the overall number of miners. Therefore, the problem of block propagation optimization is to minimize the overall AoB of block propagation with the given reputation constraint, which can be formulated as 
\begin{equation} 
\begin{split}
    &\min_{\boldsymbol{\pi}}\:\sum_{i=2}^m\Big(\Delta_i = \Gamma_i + \frac{1}{\mu}+\frac{\mu \Gamma_i^3}{1-\mu \Gamma_i}\Big)\\
    &\:\:\mathrm{s.t.}\:R_{\pi_{l}\rightarrow \pi_{l+1}}^{final}>\sigma, l\in \{1,\ldots,m-1\},\label{fu}
\end{split}
\end{equation}
where $\sigma \in(0,1)$ is a miner reputation threshold for ensuring block propagation. Considering $\Gamma_i < (1/\mu)$\cite{Akosta2017age}, it is easy to prove that the AoB $\Delta_i$ will increase as block propagation time $\Gamma_i$ increases. Moreover, it is obvious that $\Gamma_i$ is a monotonic function concerning the distance $d_i$ between miner $(i-1)$ and miner $i$ according to (\ref{t_p}). Motivated by the above analysis, we propose a GAT-based block propagation optimization model to solve the optimization problem, described in Section \ref{V}.

In this section, we propose a GAT-based block propagation optimization model for blockchain-enabled Web 3.0, which minimizes the average AoB of block propagation between adjacent miners with the specific reputation constraint, thereby obtaining the optimal block propagation trajectory. 

In public blockchains, the miner network can be regarded as a graph network\cite{wen2022optimal}. As a powerful tool that is specifically designed for processing graph-structured data\cite{Gkool2018GNNyuanxing}, we apply the GAT to find the optimal block propagation trajectory, where GAT utilizes attention mechanisms to capture the relationships between miners, enhancing the extraction of relational representations among each miner. As shown in Fig. \ref{system}, the GAT architecture of the proposed framework consists of encoder and decoder components.
More precisely, the encoder is used to extract the structural characteristics of the miner network, and the decoder systematically generates the miner sequence of the optimal block propagation trajectory. 

For the problem of block propagation optimization, $G$ as an input instance can be considered as a fully connected graph  $G = (\mathcal{M}, \mathcal{E})$, representing the public blockchain network composed of miners, where $\mathcal{E}$ is the edge set of the miner network. Considering that miner $j$ is an adjacent miner of miner $i$, we define $\boldsymbol{f}_i$ and $\boldsymbol{f}_j$ as the location features of miner $i$ and miner $j$, respectively. Specifically, the location feature $\boldsymbol{f}_i$ of miner $i$ is a $d_f$-dimensional feature, where $d_f=2$ is composed of horizontal and vertical coordinates of miner $i$ in the input instance $G$. Thus, the distance between adjacent miners can be calculated by\cite{lei2022solve}
\begin{equation}
    Dis(i,j)=\left|| \boldsymbol{f}_i-\boldsymbol{f}_j|\right|_2,\: i\neq j,\:\forall i,j\in\mathcal{M},
\end{equation}
where $\left||\cdot|\right|_2$ denotes the $L2$ norm  to calculate the $2D$ Euclidean distance. 
For network parameters $\boldsymbol{\theta}$ and the input instance $G$, the corresponding solution probability $p_{\boldsymbol{\theta}}\left(\boldsymbol{\pi}|G\right)$ is given by
\begin{equation} \label{1}
    p_{\boldsymbol{\theta}}\left({\boldsymbol{\pi}}|G\right)=\prod_{t=1}^{m}{p_{\boldsymbol{\theta}}\left({\pi}_t |G, {{\boldsymbol{\pi}}_{1:t-1}}\right)}.
\end{equation}
Therefore, the overall route length of block propagation is given by\cite{Gkool2018GNNyuanxing}
\begin{equation}
    L(\boldsymbol{\pi}|G)=\sum_{t=1}^{m-1}\left|\left| \boldsymbol f_{\pi_t}-\boldsymbol f_{\pi_{t+1}}\right|\right|_2.\label{L}
\end{equation}
The function $L(\boldsymbol{\pi}|G)$ will be utilized to train the network in Section \ref{VI}. Next, we introduce the encoder-decoder architecture constructed for the GAT-based block propagation optimization model.

\subsection{Encoder}
Unlike the typical transformer architecture\cite{vaswani2017attention}, we embed miner features into the context of the graph in the encoder\cite{Gkool2018GNNyuanxing}. The miner features are fed into a  $d_h$- dimensional hidden layer, where $d_h=128$. 
In the hidden layer, performing a learnable linear projection using parameters $\boldsymbol{W}^h$ and $\boldsymbol{b}^h$, the initial $d_h$-dimensional miner embeddings $\boldsymbol h_{i}^{(0)}$ is calculated by\cite{Gkool2018GNNyuanxing}
\begin{equation}
    \boldsymbol h_i^{(0)}=\boldsymbol{W}^h{\boldsymbol f_i}+\boldsymbol{b}^h.
\end{equation}
where $\boldsymbol{W}^h \in \Bbb{R}^{d_h\times d_x}$ and $\boldsymbol{b}^h \in \Bbb{R}^{d_h}$.
After obtaining the embedding $\boldsymbol h_{i}^{(0)}$, it is sent to the graph attention layer and updated with $N$ GAT layers. We denote $\boldsymbol h_{i}^{(r)}$ as the miner embeddings calculated by GAT layer $r\in \{1,\ldots,N\}$. Specifically, $\bar{\boldsymbol{{h}}}^{(graph)}$ is denoted as the graph embedding, which is the aggregated embedding of the input graph as the average of final miner embeddings, given by\cite{Gkool2018GNNyuanxing}
\begin{equation} \label{31}
    \bar{\boldsymbol{h}}^{(graph)}=\frac1M\sum_{i=1}^M \boldsymbol h_i^{(N)}.
\end{equation}
Finally, the encoder outputs the final miner embeddings $\boldsymbol h_i^{(N)}$, as well as the graph embedding $\bar{\boldsymbol{h}}^{(graph)}$ to the decoder.

Then, we introduce the major architecture of the proposed graph attention layer in detail. 
\subsubsection{Multi-Head Attention (MHA) layer}
The MHA layer is beneficial to learning information from different aspects than only a single-head attention mechanism\cite{Gxin2021multi-attention}. Inspired by \cite{vaswani2017attention, hwang2021GNNGAT, Gguo2020GNNGAT2}, we utilize the MHA layer to model the relevance between miners on the graph. In the MHA layer, the value\footnote{
In attention mechanisms, a query represents the focused content, a key serves as a reference, and a value corresponds to the actual information associated with the key\cite{vaswani2017attention}.} of a miner is the compatibility of the query and the key from its neighbors\cite{vaswani2017attention}. We denote the number of attention heads as $y\in\{1,\ldots, Y\}$ and consider a sequence of
query $\boldsymbol{Q}=\{\boldsymbol{q}_{1}^{(r)},\ldots,\boldsymbol{q}_{i}^{(r)},\ldots,\boldsymbol{q}_{M}^{(r)}\}$, key $\boldsymbol{K}=\{\boldsymbol{k}_{1}^{(r)},\ldots,\boldsymbol{k}_{i}^{(r)},\ldots,\boldsymbol{k}_{M}^{(r)}\}$, and value $\boldsymbol{V}=\{\boldsymbol{v}_{1}^{(r)},\ldots,\boldsymbol{v}_{i}^{(r)},\ldots,\boldsymbol{v}_{M}^{(r)}\}$\cite{Gli2019QKV}. For miner $i$, $\boldsymbol q_{i}^{(r)}$, $\boldsymbol k_{i}^{(r)}$, and $\boldsymbol v_{i}^{(r)}$ are calculated by projecting the embedding $\boldsymbol h_{i}^{(r-1)}$, which are given by
\begin{equation} \label{14}
    \boldsymbol q_{i}^{(r)}=\boldsymbol{W}^Q \boldsymbol h_{i}^{(r-1)},
\end{equation}
\begin{equation} \label{33}
    \boldsymbol k_{i}^{(r)}=\boldsymbol{W}^K\boldsymbol h_{i}^{(r-1)},
\end{equation}
\begin{equation} \label{34}
    \boldsymbol v_{i}^{(r)}=\boldsymbol{W}^V\boldsymbol h_{i}^{(r-1)},
\end{equation}
where each attention head $y$ obtains parameters $\boldsymbol{W}^{Q}\in \Bbb{R}^{d_{k}\times d_{h}}$, $\boldsymbol{W}^{K}\in \Bbb{R}^{d_{k}\times d_{h}}$, and $\boldsymbol{W}^{V}\in \Bbb{R}^{d_{v}\times d_{h}}$.

Based on (\ref{14}), (\ref{33}), and (\ref{34}), we compute the compatibility $\boldsymbol u_{ij}^{(r)}\in \Bbb{R}$ by combining the query $\boldsymbol q_{i}^{(r)}$ with the key $\boldsymbol k_{j}^{(r)}$ of miner $j$ as the dot-product function\cite{vaswani2017attention}
\begin{equation}
    \boldsymbol u_{ij}^{(r)}=\begin{cases}\frac{\left(\boldsymbol q_{i}^{(r)}\right)^T\boldsymbol k_{j}^{(r)}}{\sqrt{d_k}}&\text{if miner $i$ is adjacent to miner $j$,}\\-\infty&\text{otherwise,}\end{cases}
\end{equation}
 where the compatibility of non-adjacent miners is considered as $-\infty$, which is effective in preventing block propagation between non-adjacent miners. Then, based on the compatibilities $\boldsymbol u_{ij}^{(r)}$, the attention weight $\boldsymbol a_{ij}^{(r)}\in[0,1]$ is given by\cite{Gkool2018GNNyuanxing} 
\begin{equation} \label{16}
    \boldsymbol a_{ij}^{(r)}=softmax\left(\boldsymbol u_{ij}^{(r)}\right)=\frac{e^{\boldsymbol u_{ij}^{(r)}}}{\sum_{j=1}^Je^{\boldsymbol u_{ij}^{(r)}}},
\end{equation}
where $\{1,\ldots,j,\ldots,J\}\subset \mathcal{M}$ is the adjacent miner set of miner $i$. Then, based on (\ref{34}) and (\ref{16}), the result vector $\boldsymbol h_{i}^{\prime}(r)$, combining $\boldsymbol a_{ij}^{(r)}$ with $\boldsymbol v_{j}^{(r)}$, is expressed as\cite{Gkool2018GNNyuanxing} 
\begin{equation} \label{17}
    \boldsymbol h_{i}^{\prime}{}^{(r)}=\sum_{j=1}^J \boldsymbol a_{ij}^{(r)}\boldsymbol v_{j}^{(r)},
\end{equation}
Finally, in the MHA layer, miners are allowed to receive different types of information from their neighbors to transform the parameters $\boldsymbol Q$, $\boldsymbol K$, and $\boldsymbol V$ into different and learnable projections\cite{lei2022solve}. Specifically, we use $Y=8$ heads and  $d_k=d_v=d_h/Y=16$. Besides, we define $\boldsymbol{W}_y^{O^{(r)}}\in \Bbb{R}^{d_h\times d_v}$, and the final value of the MHA layer for miner $i$ in the graph attention layer $r$ is projected to a single $d_h$-dimensional vector, given by
\begin{equation}\label{18}
    \\{MHA}_i^{(r)}\left(\boldsymbol {h}_1^{(r)},\ldots,\boldsymbol h_M^{(r)}\right)=\sum_{y=1}^Y\boldsymbol{W}_y^{O^{(r)}}\boldsymbol h_{iy}^{\prime}{}^{(r)}.
\end{equation}
\subsubsection{Batch Normalization (BN) layer}

    In the BN layer, we introduce learnable parameters  $\boldsymbol{w}^{batch}$ and $\boldsymbol{b}^{batch}$ as the $d_h$-dimensional affine parameters, and $\overline{BN}^{(r)} (\boldsymbol h_i^{(r)})$ is denoted as batch normalization without affine transformation. Besides, we use $\odot$ to represent the element-wise product. Therefore, $BN^{(r)}(\boldsymbol h_i^{(r)})$ is expressed as\cite{Gkool2018GNNyuanxing} 
\begin{equation}
    BN^{(r)}(\boldsymbol h_i^{(r)})=\boldsymbol{w}^{batch}\odot\overline{BN}^{(r)}\left(\boldsymbol h_i^{(r)}\right)+\boldsymbol{b}^{batch}.
\end{equation}
\subsubsection{Feed-Forward (FF) layer}

In the FF layer, we use a  $d_{F}$-dimensional hidden sublayer, where $d_{F}=512$, with the parameters $\boldsymbol{W}^{f2}$ and $\boldsymbol{b}^{f2}$ and a $ReLu$ activation with the parameters $\boldsymbol{W}^{f1}$ and $\boldsymbol{b}^{f1}$ to construct the FF layer:
\begin{equation}
    FF^{(r)}\left(\hat{\boldsymbol {h}}_i^{(r)}\right)=\boldsymbol{W}^{f2}\cdot ReLu\left(\boldsymbol{W}^{f1} \hat{\boldsymbol{{h}}}_i^{(r)}+\boldsymbol{b}^{f1}\right)+\boldsymbol{b}^{f2},
\end{equation}
where the input of the FF layer $\hat{\boldsymbol {h}}_i^{(r)}$ is the output of the BN layer after the MHA layer, as calculated in (\ref{hat h}).

\subsubsection{Graph attention network layer}

Based on the proposed three key layers, i.e., the MHA layer, the BN layer, and the FF layer, the GAT layer is given by\cite{Gxin2021multi-attention}

\begin{align}
    \boldsymbol{\hat{h}}_i^{(r)} &= BN^{(r)}\left(\boldsymbol h_i^{(r-1)}+MHA_i^{(r)}\left(\boldsymbol h_1^{(r-1)},\ldots,\boldsymbol h_M^{(r-1)}\right)\right),\label{hat h} \\
    \boldsymbol{h}_i^{(r)} &= BN^{(r)}\left(\boldsymbol{\hat  h}_i^{(r)}+FF^{(r)}\left(\boldsymbol{\hat h}_i^{(r)}\right)\right),\label{42}
\end{align}
where the layers are connected by a skip-connection\cite{Gkool2018GNNyuanxing}. Then, the result of (\ref{42}) in layer $N$ will be input to (\ref{31}) to generate the aggregated $\bar{\boldsymbol{h}}^{(graph)}$.

\subsection{Decoder}
 The decoding step for block propagation is sequential. For each decoder step $t\in\{1,\ldots,m\}$, the current miner that would propagate the block selects its adjacent miner to append to the end of the sequential solution. We construct a special context vector $\boldsymbol{h}^{(d)}$ to represent the decoding context\cite{Gkool2018GNNyuanxing}, which is composed of the outputs of the encoder and decoder up to time $t$. According to \cite{vaswani2017attention}, $\boldsymbol{h}^{(d)}$ is given by
\begin{equation}
    \boldsymbol{h}^{(d)}=\begin{cases}\left[\bar{\boldsymbol{h}}^{(graph)},\boldsymbol{v}^1,\boldsymbol{v}^2\right]&\quad \text{if}\quad{t}=1,\\\left[\bar{\boldsymbol{h}}^{(graph)},\boldsymbol{v}^2,\boldsymbol{h}_{{\pi}_1}^{(N)}\right]&\quad \text{if}\quad{t}=2,\\\left[\bar{\boldsymbol{h}}^{(graph)},\boldsymbol{h}_{{\pi}_{t-2}}^{(N)},\boldsymbol{h}_{{\pi}_{t-1}}^{(N)}\right]&\quad\text{if}\quad{t}>2. \end{cases}
\end{equation}
Here, $[\cdot,\cdot,\cdot,]$ is the horizontal concatenation operator. For each time $t$, we utilize the graph embedding $\bar{\boldsymbol{h}}^{(graph)}$ as a part of the context vector $\boldsymbol{h}^{(d)}$, for the reason that the graph embedding is designed for taking the complete graph structure into account. For $t=1$, we introduce learnable $d_h$-dimensional parameters $\boldsymbol v^1$ and $\boldsymbol v^2$ as input placeholders. For $t=2$, the placeholder $\boldsymbol v_1$ would be replaced by the embedding $\boldsymbol h_{\pi_{1}}^{(N)}$. For $t>2$, the placeholder $\boldsymbol v_2$ and the embedding $\boldsymbol h_{\pi_{1}}^{(N)}$ would be replaced by the embeddings $\boldsymbol h_{\pi_{t-1}}^{(N)}$ and $\boldsymbol h_{\pi_{t-2}}^{(N)}$ of miners $\pi_{t-1}$ and $\pi_{t-2}$, to achieve the local optimum of AoB, and eventually achieve the optimization objective of minimizing overall AoB. Note that the result vector $\boldsymbol h^{(d)}$ is $(3\cdot d_h)$-dimension to align with the miner embeddings $\boldsymbol h_{\pi_t}^{(N)}$.
\begin{algorithm}[t] \label{al}
    \caption{Reinforcement Learning-based Graph Attention Network for Block Propagation Optimization}
    \KwIn{Basic physical parameters $\small\{B_{block}, \rho_s, c^0, \varepsilon, W, N_0\small\}$, batch size $B$, number of epochs $E$, steps per epoch $T$, significance of  the paired t-test $\alpha^{gat}$, miner reputation threshold $\sigma$, number of miners $M$.}
    \For{${i}= 1, \ldots, M$}
            {   
                Record the miner interactions $\alpha^{t_k}$ and $\beta^{t_k}$. 
                
                Calculate the miner reputation $R_{i\to j}^{final}$.
            }
    
        Initialize $\boldsymbol\theta$, $\boldsymbol\theta^{BL}$ $\gets$ $\boldsymbol\theta$.
        
    \For{$epoch = 1,\ldots,E$}
        {   \For{$step = 1,\ldots,T$}
            {
                $x_i\gets$ RandomInstance( )\cite{Gkool2018GNNyuanxing}, $\forall{i}\in\{1, \ldots, B\}$.

            \For{${i}= 1, \ldots, M$}
            {   
                \If{$R_{i\to j}^{final}$ < $\sigma$}
            {
                Mask miner $i$ to avoid the low-reputation miners.
            } 
            }

                $\boldsymbol\pi_i\gets$ SampleRollout($G_i, \boldsymbol p_{\boldsymbol\theta}$)\cite{Gkool2018GNNyuanxing}, $\forall{i}\in\{1, \ldots, B\}$.
    
                $\boldsymbol\pi_i^{BL}\gets$ GreedyRollout($G_i, p_{\boldsymbol\theta}^{BL}$)\cite{Gkool2018GNNyuanxing}, $\forall{i}\in\{1, \ldots, B\}$.
    
                $\nabla\mathcal{L}\leftarrow\sum_{i=1}^{B}\left(L(\boldsymbol{\pi_{i}|G)})-L(\boldsymbol{\pi_{i}^\mathrm{BL}|G)})\right)\nabla_{\boldsymbol{\theta}}\log p_{\boldsymbol{\theta}}(\boldsymbol{\pi_{i}})$.
                
                $\boldsymbol\theta \gets$ Adam($\boldsymbol\theta, \nabla\mathcal{L}$).
            }
            
            \If {OneSidedPairedTTest($p_{\boldsymbol\theta}, p_{\boldsymbol\theta}^{BL}$) < $\alpha^{gat}$\cite{Gkool2018GNNyuanxing}}
            {
                $\boldsymbol\theta^{BL}\gets\boldsymbol\theta$.
            }

        }
   
    Feed the miners with their reputations into the trained network.
    
    Calculate the minimized overall AoB and the total value of miner reputations.
    
\KwOut {The optimal block propagation trajectory, the minimized overall AoB, and the total value of miner reputations.}
\end{algorithm}
In the decoder, we compute compatibility using an MHA layer similar to the encoder. Firstly, the aggregated query $\boldsymbol q^{(d)}$ is calculated by the context embedding $\boldsymbol h^{(d)}$ while the keys $\boldsymbol k_i$ and the values $\boldsymbol v_i$ are calculated by the miner embeddings $\boldsymbol h_i^{(N)}$\cite{Gkool2018GNNyuanxing}:
\begin{align}
    \boldsymbol{q}^{(d)} & = \boldsymbol{W}^Q\boldsymbol{h}^{(d)}, \label{35} \\
    \boldsymbol{k}_i & = \boldsymbol{W}^K\boldsymbol{h}_i^{(N)}, \\
    \boldsymbol{v}_i & = \boldsymbol{W}^V\boldsymbol{h}_i^{(N)}.
\end{align}

Based on (\ref{35}), we can use the single query $\boldsymbol{q}^{(d)}$ to compute its compatibility with all miners. If miners have already received the block at time $t$, we will set the compatibility to $-\infty$, masking miners that have already received blocks to avoid selecting them repeatedly. Therefore, the $u_j^{(d)}$ is given by\cite{Gkool2018GNNyuanxing}
\begin{equation}
    u_j^{(d)}=\begin{cases}\frac{\boldsymbol q^{(d)T}\boldsymbol k_j}{\sqrt{d_k}}&\quad \text{if}\quad j\neq\pi_{t^{\prime}},\forall t^{\prime}<t,\\-\infty&\quad \text{otherwise},\end{cases}
\end{equation}
where ${\boldsymbol q^{(d)T}}$ denotes the transpose of ${\boldsymbol q^{(d)}}$. Then, we update the final context embedding $\boldsymbol{h}^{(d)}$ and miner embeddings $\boldsymbol{h}_i^{(N)}$ based on (\ref{16}), (\ref{17}), and (\ref{18}). By using a single-head attention layer, which means that we set $Y=1$, the compatibilities $u_j^{(d)}$ is calculated by\cite{Gkool2018GNNyuanxing}
\begin{equation}
    u_j^{(d)}=\begin{cases}C\cdot\tanh\Big(\frac{\mathbf{q}^{(d)T}\mathbf{k}_j}{\sqrt{d_k}}\Big)&\text{if}\quad j\neq\pi_{t^{\prime}},\forall t^{\prime}<t,\\-\infty&\text{otherwise,} \end{cases}
\end{equation}
where we clip the function $\tanh(\cdot)$ within $[-C,C]$ and $C = 10$\cite{Gkool2018GNNyuanxing}. In the last step, by using the softmax, the probability of miner $i$ chosen to propagate the block is given by\cite{Gkool2018GNNyuanxing}
\begin{equation}
    p_i=p_{\boldsymbol{\theta}}(\pi_t=i|G,\boldsymbol{\pi}_{1:t-1})=\frac{e^{u_j^{(d)}}}{\Sigma_{j=1}^Je^{u_j^{(d)}}}.
\end{equation}
Finally, we can obtain the solution path of the optimal block propagation trajectory based on (\ref{1}).

\begin{figure*}[t]
\centering
\subfigure[Channel bandwidth $b=180kHz$.]
{
    \begin{minipage}[t]{0.31\linewidth}
	\centering
	\includegraphics[width=1.1\linewidth]{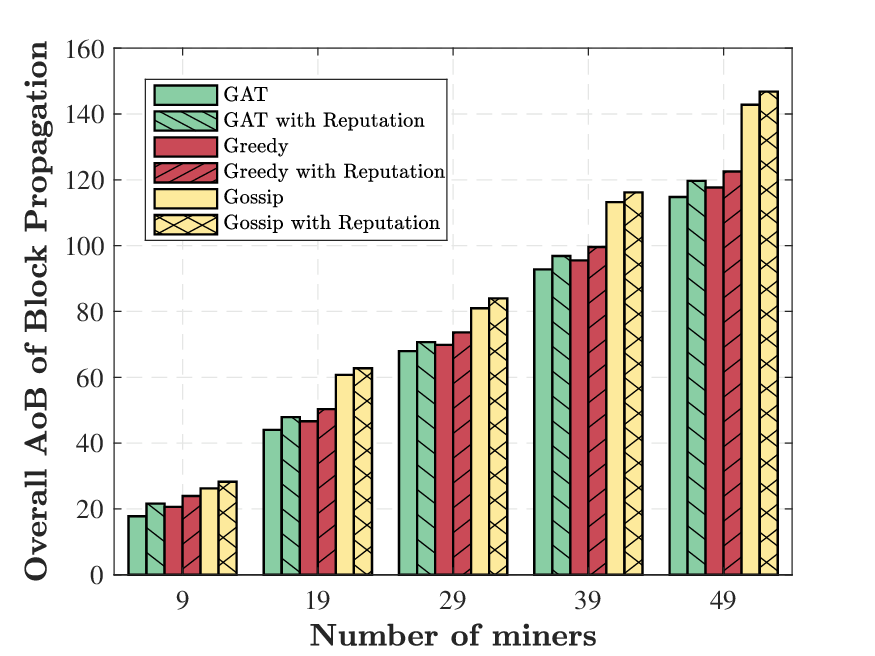}
        \captionsetup{font=footnotesize}
    \end{minipage}
}
\subfigure[Channel bandwidth $b=22MHz$.]
{
    \begin{minipage}[t]{0.31\linewidth}
	\centering
	\includegraphics[width=1.1\linewidth]{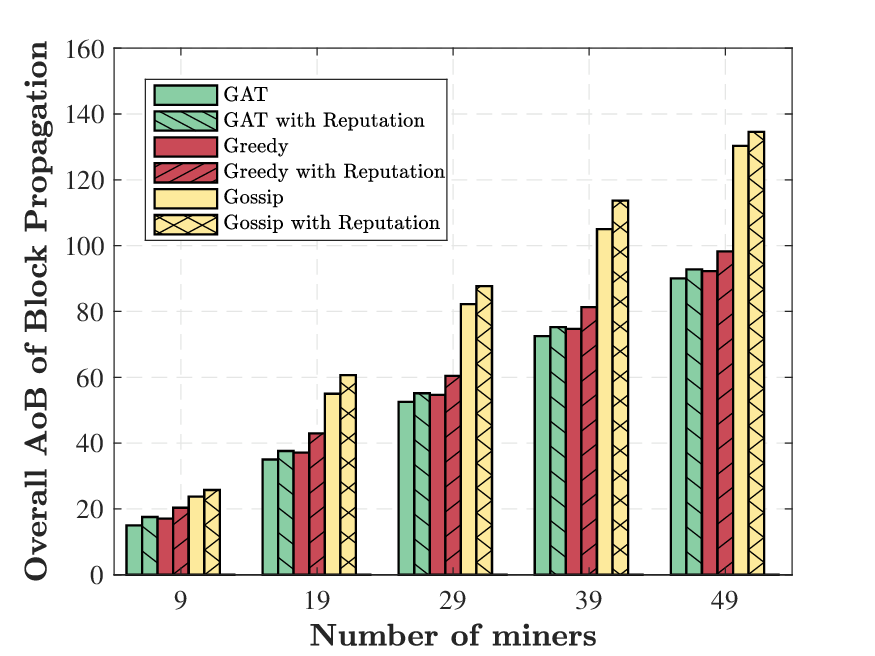}
        \captionsetup{font=footnotesize}
    \end{minipage}
}
\subfigure[Channel bandwidth $b=100MHz$.]
{
    \begin{minipage}[t]{0.31\linewidth}
	\centering
	\includegraphics[width=1.1\linewidth]{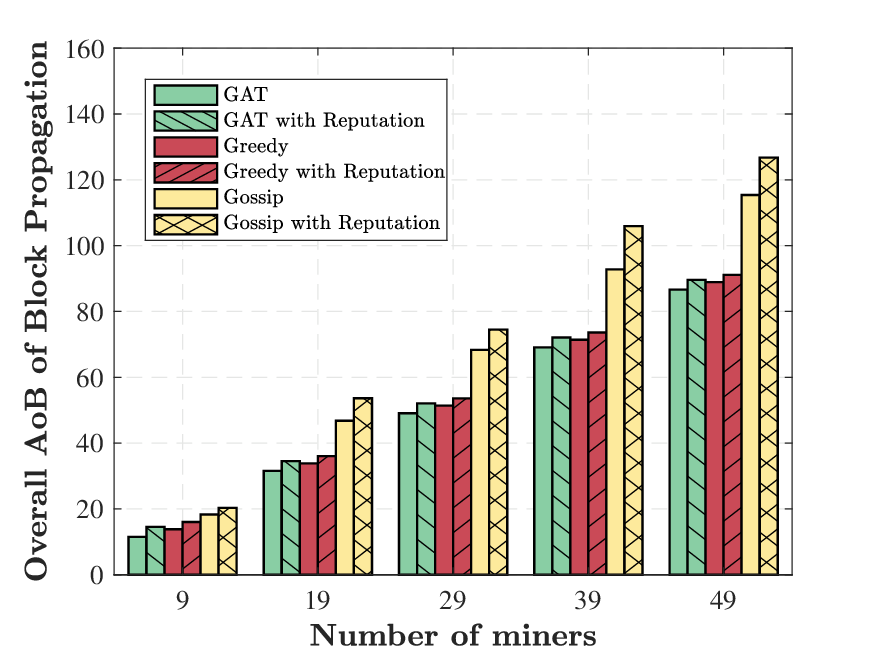}
        \captionsetup{font=footnotesize}
    \end{minipage}
}
\caption{The overall AoB of block propagation corresponding to different numbers of miners in different channel bandwidths.}\label{AoBendd}
\end{figure*}

\section{Model Training}\label{VI}
In this section, we first propose an algorithm to solve the block propagation optimization problem, as shown in \textbf{Algorithm 1}. In the beginning, we record the miner transaction in a period $t_k$ to calculate the miner reputation $R_{i\to j}^{final}$, and then we train the GAT model. Based on (\ref{L}), the loss of the model is the expectation of the cost $L(\boldsymbol{\pi}|G)$\cite{Gkool2018GNNyuanxing}, given by
 \begin{equation}
     \mathcal{L}(\boldsymbol{\theta}|G)=\mathbb{E}_{p_{\boldsymbol{\theta}}(\boldsymbol{\pi}|G)}[L(\boldsymbol{\pi}|G)].
 \end{equation}
 We optimize $\mathcal{L}(\boldsymbol{\theta}|G)$ based on the gradient descent, employing a gradient estimator with baseline $b(G)$, which is given by\cite{Gkool2018GNNyuanxing}
 \begin{equation}
     \nabla\mathcal{L}(\boldsymbol{\theta}|G)=\mathbb{E}_{p_{\boldsymbol\theta}(\boldsymbol{\pi}|G)}\left[\left(L(\boldsymbol{\pi}|G)-b(G)\right)\nabla\log p_{\boldsymbol{\theta}}(\boldsymbol{\pi}|G)\right],
 \end{equation}
where $b(G)$ denotes the baseline utilized in the network for effective training.  Fundamentally, a baseline serves the purpose of estimating the complexity of a given instance $G$, allowing for a correlation with the cost $L(\boldsymbol{\pi}|G)$ to evaluate the advantage of the model-selected solution $\boldsymbol{\pi}$. Therefore, we incorporate the rollout baseline into our training process.

\begin{figure}[t]
    \begin{center}

    \centering
    \includegraphics[width=1\linewidth]{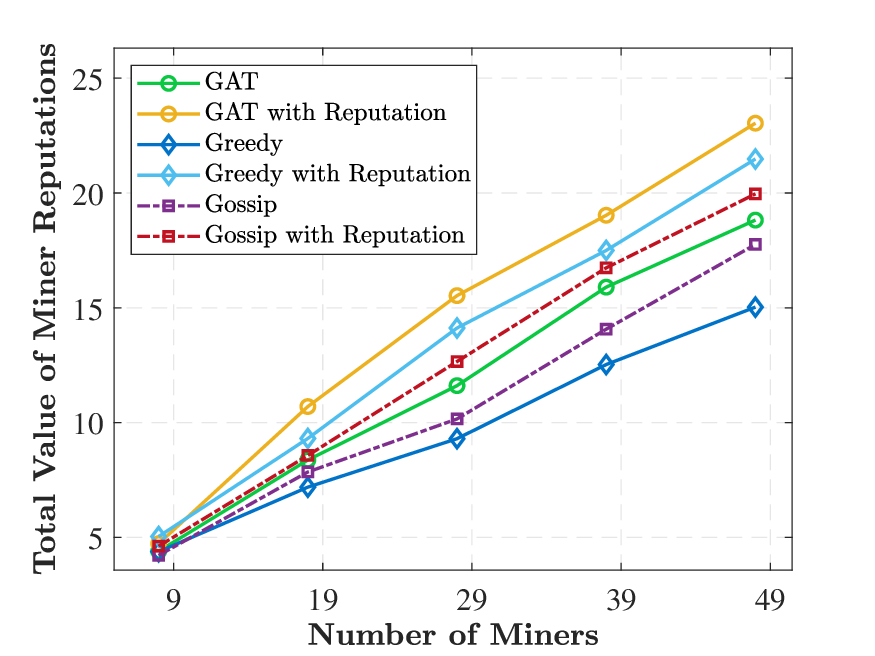}
    \caption{The total value of miner reputations corresponding to different numbers of miners.}\label{repution}  

\end{center}
\end{figure}
\begin{table}[t]
	\renewcommand{\arraystretch}{1.4} 
	\caption{ Key Parameters in the Simulation.}\label{table} \centering 
	\begin{tabular}{m{4.7cm}<{\raggedright}|m{2.7cm}<{\centering}} 
		\hline		
		\textbf{Parameters} & \textbf{Values}\\		
  	\hline
		Size of block $(B_{block})$ & $1$\:$\rm{MB}$  \\
		\hline
		Transmit power of the IoT devices $(\rho_s)$ & $23$\:$\rm{dBm}$  \\	
		\hline
		Noise power density $(N_0)$ & $-174$\:$\rm{dBm/Hz}$  \\
		\hline
		Path-loss coefficient $(\varepsilon)$  &  $3.38$  \\	
		\hline		
		Channel  bandwidth between adjacent miners $(b)$ &  $180$\:$\rm{kHz}$, $22$\:$\rm{MHz}$, and $100$\:$\rm{MHz}$ \\
		\hline	
            Unit channel gain $(c^0)$ &  $-30$\:$\rm{dB}$ \\
        \hline	
	\end{tabular}\label{parameter}
\end{table}
In line with the self-critical training introduced by \cite{grennie2017self}, our approach involves periodic updates of the network parameters $\boldsymbol\theta^{BL}$, which are specifically for the baseline policy. In detail, we conduct a paired t-test with a significance level of $\alpha^{gat} = 5\%$ to compare the baseline policy with the newest training policy upon completion of each epoch on the separated testing instances. If the improvement is statistically significant, the baseline parameters $\boldsymbol\theta^{BL}$ will be updated by the network parameters $\boldsymbol\theta$\cite{Gkool2018GNNyuanxing}. Conducting evaluations upon completion of each epoch ensures that the current model consistently faces challenges from the best possible model available. Moreover, to eliminate variance, we enforce determinism by greedily selecting actions based on their maximum probabilities. Moreover, by utilizing the greedy rollout as our baseline $b(G)$, the function $L(\boldsymbol{\pi}|G) - b(G)$ yields a negative value when the sampled solution $\boldsymbol \pi$ surpasses the performance of the greedy rollout based on the paired t-test, thereby resulting in the reinforce of actions, and vice versa\cite{Gkool2018GNNyuanxing}. 
\begin{figure*}[t]
\centering
\subfigure[The block propagation trajectory in 9 miners, which obtains $\Delta =21.61$ and total value of miner reputation $R_{total}=4.40$.]
{
    \begin{minipage}[t]{0.3\linewidth}
	\centering
	\includegraphics[width=1.1\linewidth]{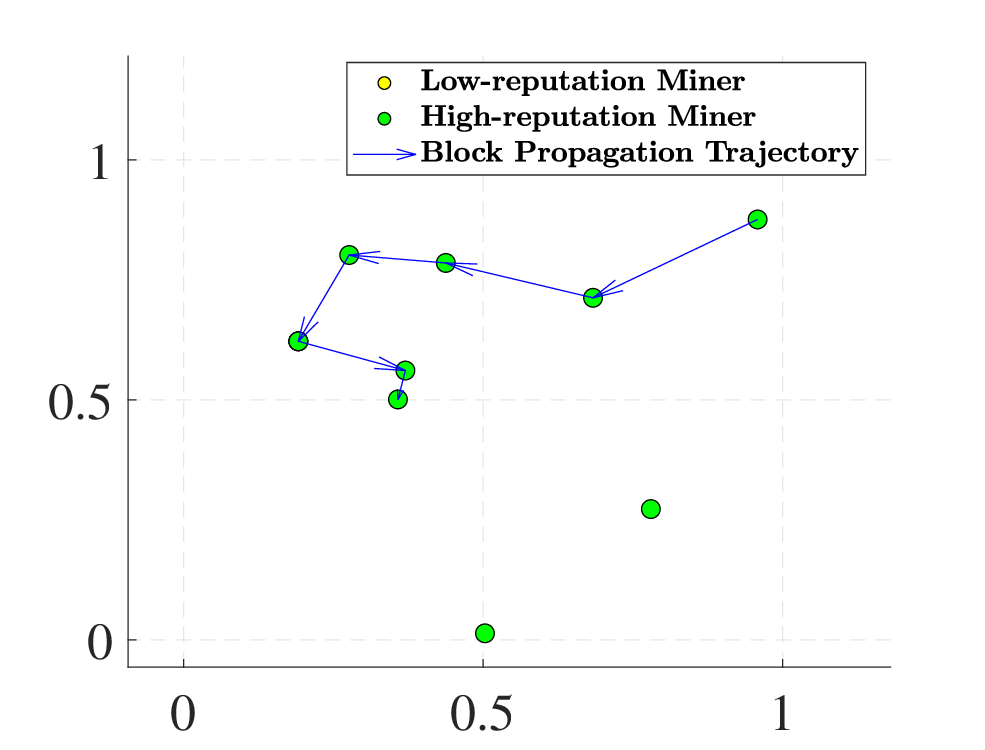}
        \captionsetup{font=footnotesize}
	\label{9}
    \end{minipage}
}
\subfigure[The block propagation trajectory in 19 miners, which obtains $\Delta =47.87$ and total value of miner reputation $R_{total}=10.70$.]
{
    \begin{minipage}[t]{0.3\linewidth}
	\centering
	\includegraphics[width=1.1\linewidth]{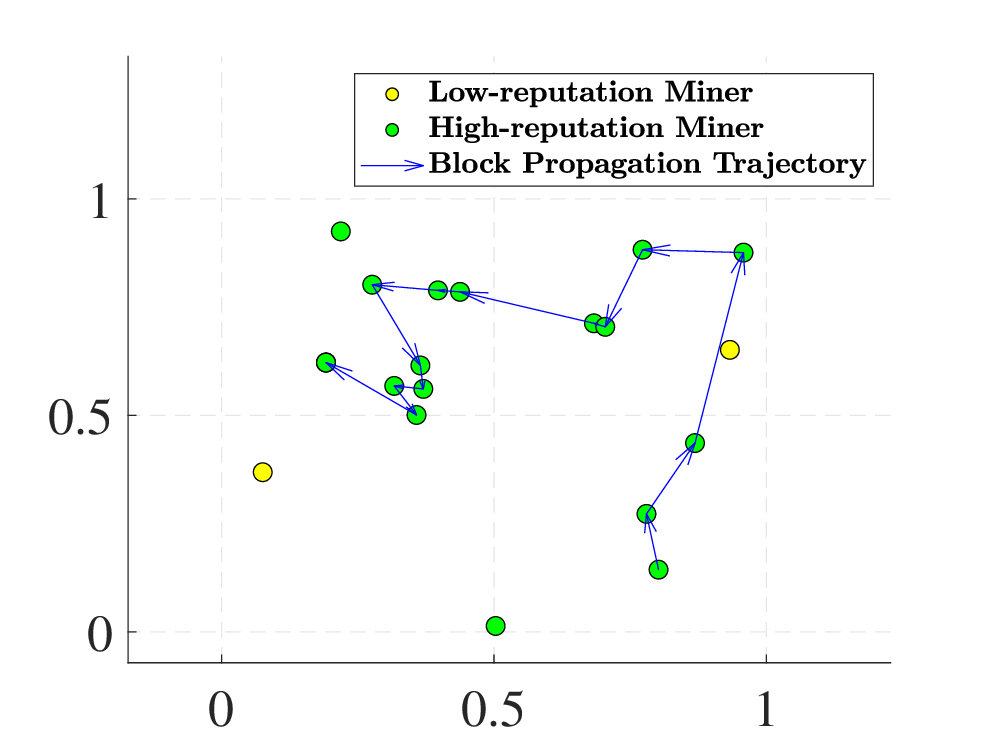}
        \captionsetup{font=footnotesize}
	\label{19}
    \end{minipage}
}
\subfigure[The block propagation trajectory in 29 miners, which obtains $\Delta =70.69$ and total value of miner reputation $R_{total}=15.53$.]
{
    \begin{minipage}[t]{0.3\linewidth}
	\centering
	\includegraphics[width=1.1\linewidth]{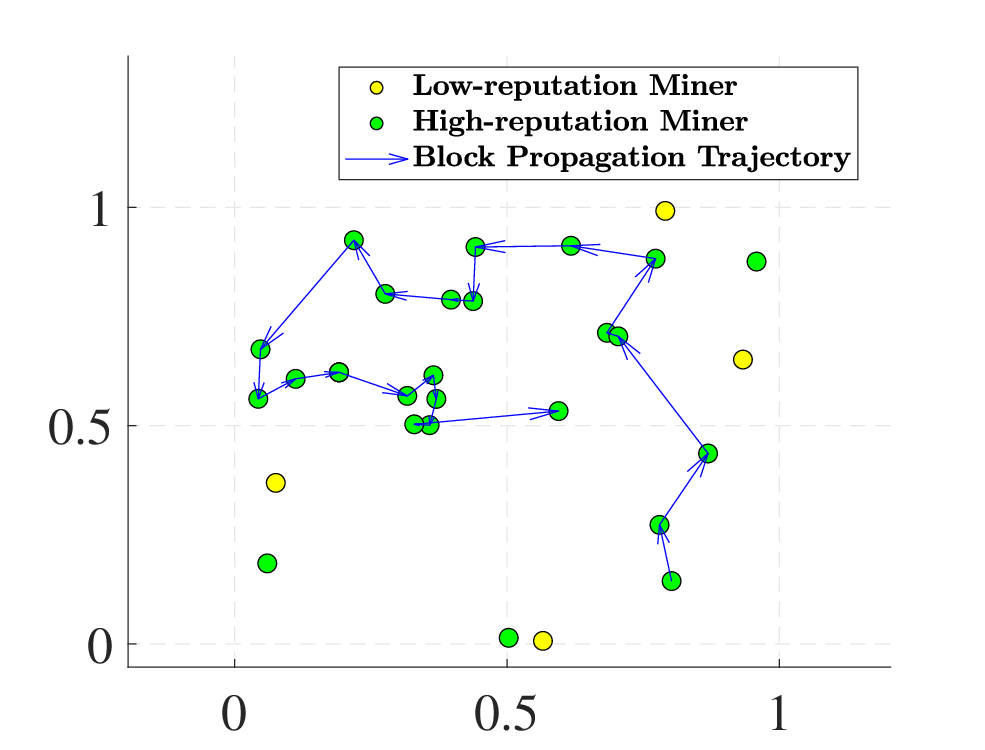}
        \captionsetup{font=footnotesize}
	\label{29}
    \end{minipage}
}
\subfigure[The block propagation trajectory in 39 miners, which obtains $\Delta =96.84$ and total value of miner reputation $R_{total}=19.03$.]
{
    \begin{minipage}[t]{0.3\linewidth}
	\centering
	\includegraphics[width=1.1\linewidth]{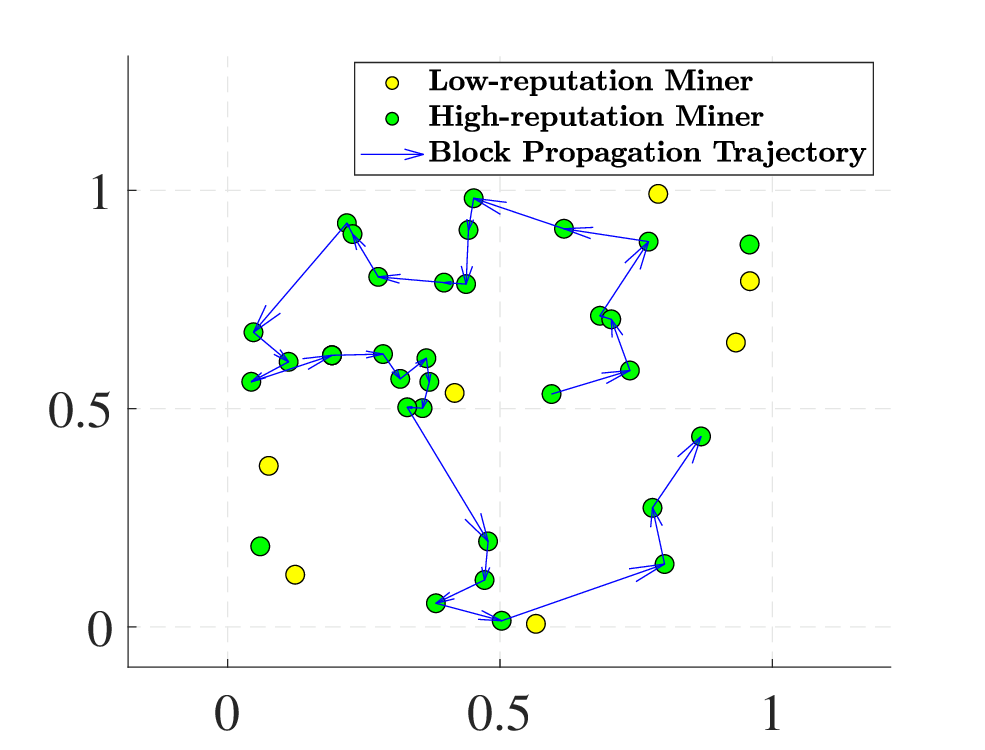}
        \captionsetup{font=footnotesize}
	\label{39}
    \end{minipage}
}
\subfigure[The block propagation trajectory in 49 miners, which obtains $\Delta =119.69$ and total value of miner reputation $R_{total}=23.04$.]
{
    \begin{minipage}[t]{0.3\linewidth}
	\centering
	\includegraphics[width=1.1\linewidth]{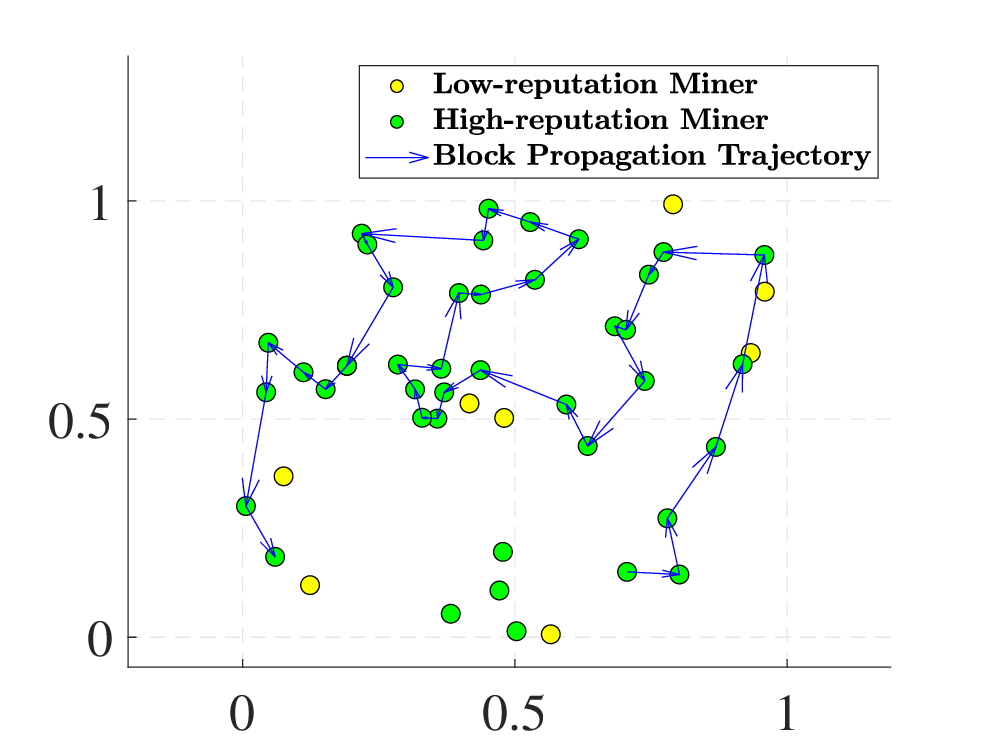}
        \captionsetup{font=footnotesize}
	\label{49}
    \end{minipage}
}
\caption{The block propagation trajectories corresponding to different numbers of miners.}\label{trajectory}
\end{figure*}
Furthermore, concerning the algorithm itself, each rollout incurs an additional forward pass, resulting in a $50\%$ increase in computational requirements\cite{Gkool2018GNNyuanxing}. However, given that the baseline policy remains constant throughout an entire epoch, we can streamline the process of sampling data and computing baselines by employing larger batch sizes. This efficiency is possible due to reduced memory requirements, allowing computations to operate in pure inference mode \cite{Gkool2018GNNyuanxing}. Note that the computational complexity of \textbf{Algorithm 1} is $O(M)$\cite{mdrori2020learning}, indicating that \textbf{Algorithm 1} is efficient. Finally, after training the network, the miners, along with their reputations, are input into the trained
network. The network minimizes the overall AoB of block propagation and calculates the total reputation of miners, generating the optimal block propagation trajectory for a specific number of miners as the final output.

\section{Numerical Results}\label{VII}

In this section, we first compare the proposed GAT model with other conventional routing mechanisms: i) \textit{Greedy mechanism}. In the simulation settings, the greedy mechanism always calculates the distance between miners and consistently selects the nearest adjacent miner to propagate the block; ii) \textit{Gossip mechanism}. In the simulation settings, the gossip mechanism only takes into account adjacent miners, and the miner randomly propagates the block to their adjacent miner. At the same time, we separately add the reputation constraints to the three proposed mechanisms. Then, we present the optimal block propagation trajectories under different miner numbers in the proposed GAT model. Finally, we evaluate the impact of baselines and learning rate settings on the performance of the proposed GAT model.  

For the parameter settings of the experiment, the parameters $\boldsymbol{\theta}$ of the GAT network are initialized uniformly within the range $\big(-1/\sqrt{d_f}, 1/\sqrt{d_f}\big)$, where $d_f=2$ represents the input dimension of miners. Moreover, we use $N = 3$ layers in the encoder, which is a thoughtful balance between the quality of results and computational complexity. We set $r^{ratio}=3/4$ to ensure that the block propagates to miners over $50\%$ for validation. Additionally, the training spans $100$ epochs using dynamically generated training data, processing $2500$ batches of $512$ instances per epoch. Note that we run the experiments on NVIDIA GeForce RTX 3080Ti, and the main parameters of this paper are listed in Table \ref{parameter}\cite{Padhikary2016performance,  akpakwu2017survey, garcia2021tutorial, wen2023task, su2022energy}.
\begin{figure*}[t]
\centering
\subfigure[Miner numbers $M=19$.]
{
    \begin{minipage}[t]{0.45\linewidth}
	\centering
	\includegraphics[width=1.1\linewidth]{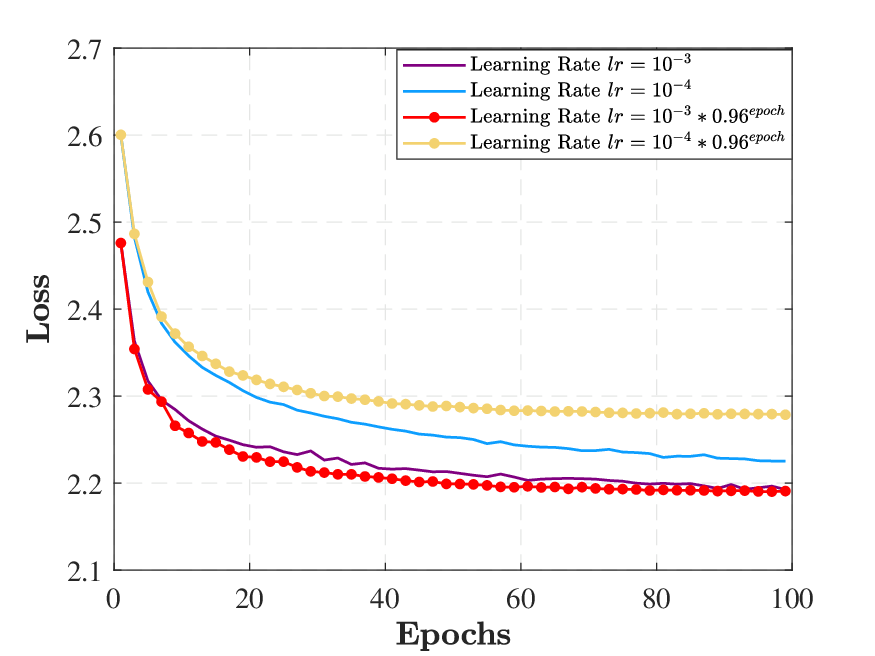}
        \captionsetup{font=footnotesize}
	\label{lr19}
    \end{minipage}
}
\subfigure[Miner numbers $M=49$.]
{
    \begin{minipage}[t]{0.45\linewidth}
	\centering
	\includegraphics[width=1.1\linewidth]{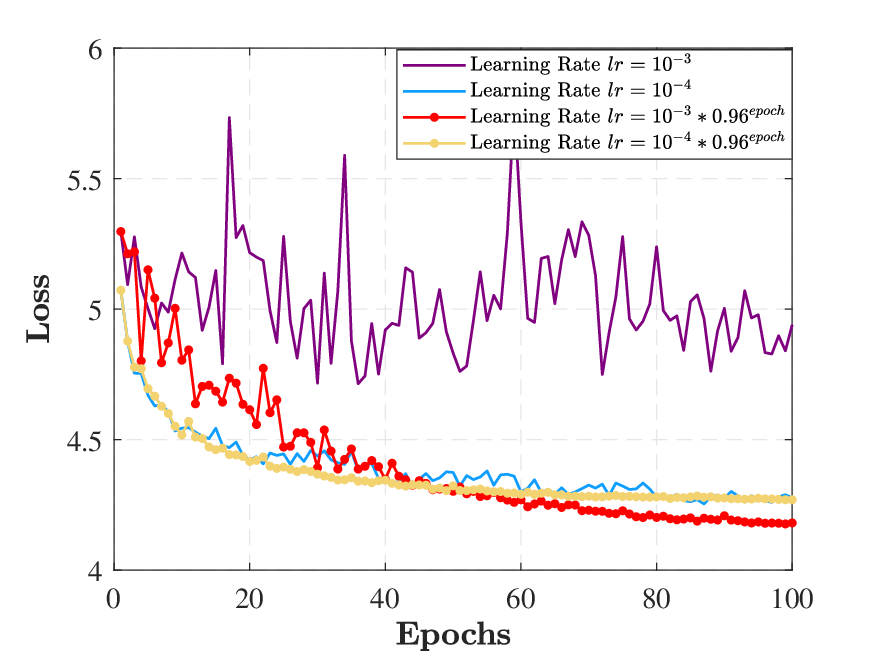}
        \captionsetup{font=footnotesize}
	\label{lr49}
    \end{minipage}
}
\caption{The network loss corresponding to the training epochs utilizing four learning rate schemes at $M=19$ and $M=49$.}\label{lr}
\end{figure*}
 \begin{figure}[t]\centering     
 \includegraphics[width=0.45\textwidth]{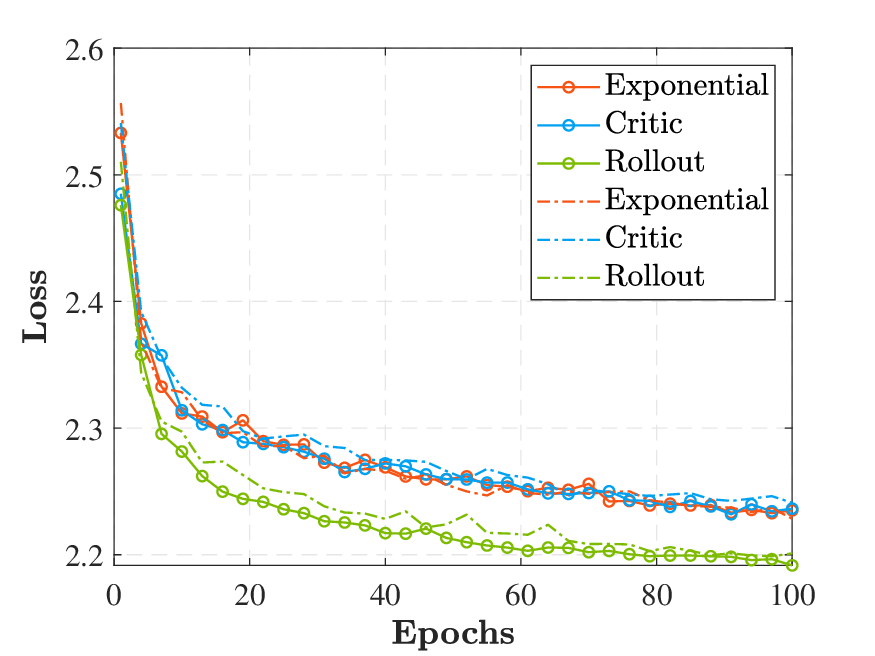}     \caption{The network loss as a function corresponding to the training epochs utilizing different baselines.}  \label{baseline}     
\end{figure}

Fig. \ref{AoBendd} shows the overall AoB of block propagation corresponding to different numbers of miners in different channel bandwidths utilizing different routing mechanisms. For convenience, we denote "GAT with Reputation Mechanism" to represent the GAT model with the reputation mechanism, which is the same as the "Greedy with Reputation Mechanism" and "Gossip with Reputation Mechanism". Considering different communication techniques, we conduct simulations with bandwidths of $b=180$\:$\rm{kHz}$\cite{Padhikary2016performance}, $b=22$\:$\rm{MHz}$\cite{akpakwu2017survey}, and $b=100$\:$\rm{MHz}$\cite{garcia2021tutorial}, corresponding to typical channels, WiFi, and 5G, respectively. From Fig. \ref{AoBendd}, we can observe that, as the channel bandwidth increases, the AoB decreases significantly, highlighting the necessity of improving channel bandwidth. Moreover, we can observe that both the proposed GAT model and the GAT model with the reputation mechanism consistently outperform the other four mechanisms. When $M=49$, the advantage is particularly pronounced, where the AoB of the GAT model is $2.5\%$ lower than that of the Greedy mechanism and $24.4\%$ lower than that of the Gossip mechanism, while the AoB of the GAT model with the reputation mechanism is $2.4\%$ lower than the Greedy mechanism with the reputation mechanism and $22.6\%$ lower than that of the Gossip mechanism with the reputation mechanism. The reason is that the GAT model focuses on minimizing the overall AoB of block propagation by considering the global structure of the miner network, which can obtain the optimal block propagation trajectory, thereby achieving block propagation optimization. In contrast, the Greedy mechanism only focuses on the selection of the current step and ignores the global structure, while the Gossip mechanism randomly selects miners without optimizing AoB, resulting in a much lower effect than other mechanisms.  

Fig. \ref{repution} depicts the total value of miner reputations corresponding to different numbers of miners under different routing mechanisms, displaying a monotonic rise with an increase in the number of miners. As we can see, the algorithms with the miner reputation mechanism can obtain the propagation trajectory with a high total of value miner reputations, thus effectively ensuring reliability in block propagation. Moreover, the GAT model with the reputation mechanism performs best among other mechanisms and the reputation gap becomes more obvious with a larger number of miners. For example, comparing the Greedy mechanism with the GAT model with the reputation mechanism, the reputation of the GAT model with the reputation mechanism is $7.4\%$ higher than that of the Greedy mechanism at $M=9$, which outstandingly increases to $34.77\%$ at $M=49$. Combining Fig. \ref{AoBendd} and Fig. \ref{repution}, we can conclude that algorithms with the reputation mechanism inevitably improve the reliability of block propagation at the cost of increasing AoB \cite{zhong2023blockchain}. However, the improvement in the reliability of block propagation with a relatively small increase in AoB is reasonable. Overall, the GAT model with the reputation mechanism stands out as the most exceptional performer among other mechanisms.

In Fig. \ref{trajectory}, we present the optimal block propagation trajectory corresponding to different miner numbers $M=\{9,19,29,39,49\}$. For better distinction, we draw yellow points to denote the low-reputation miners and green points to denote the high-reputation miners, where the reputation values are calculated by the proposed reputation mechanism\cite{Rkang2019toward}. Moreover, the blue arrows point to the optimal block propagation trajectories generated by the proposed GAT model. We can observe that the proposed GAT model can obtain optimal block propagation trajectories based on different numbers of miners, which are clearly organized and have no unreasonable costs in AoB for efficiency. At the same time, the block propagation trajectories can perfectly prevent the blocks from propagating to the low-reputation miners, which ensures the reliability of block propagation.

Fig. \ref{lr} depicts the network loss corresponding to training epochs with four learning rate schemes for miner numbers $M=19$ and $M=49$. As shown in Fig. \ref{lr19}, the GAT model with the learning rate $lr=10^{-3}$ outperforms the GAT model with the learning rate $lr=10^{-4}$, with marginal improvement from learning rate decay\cite{Gkool2018GNNyuanxing}. However, in Fig. \ref{lr49}, the loss of the GAT model with the learning rate $lr=10^{-3}$ converges slowly and fails to fully converge after $100$ epochs with a larger loss. Conversely, employing a learning rate of $lr=10^{-3}$ with a decay of $0.96^\text{epoch}$ yields better model performance, and further improvement may be achievable with prolonged training. Furthermore, motivated by the above analysis, we conclude that employing a higher learning rate of $lr=10^{-3}$ accelerates the initial learning but necessitates decay for convergence\cite{Gkool2018GNNyuanxing}, especially in the case of a larger number of miners. Moreover, a smaller learning rate of $lr=10^{-4}$ shows increased stability without the need for decay, albeit with a limited impact on loss reduction.

In Fig. \ref{baseline}, we compare the proposed rollout baseline with two other baseline schemes, i.e., the exponential baseline and the critic baseline\footnote{The encoder in critic network is similar to the GAT model, but has differences in the decoder, which has a Multilayer Perceptron characterized by a single hidden layer\cite{grennie2017self}.}. Specifically, the exponential baseline has an exponential factor of $0.8$. The experiments for baseline comparison are conducted under the miner number $M=19$ and the learning rate $lr=10^{-3}$. To demonstrate generalization across different graph data, we employ two seeds. As shown in Fig. \ref{baseline}, the solid line corresponds to $\text{seed}=1234$, and the dash-dot line corresponds to $\text{seed}=1000$. We can see that all three baselines converge effectively, and the rollout baseline exhibits the best performance with reaching the smallest loss, while the exponential and critic baselines perform closely. Moreover, the different seeds perform similarly, ensuring model generalization across diverse graph data\cite{Gkool2018GNNyuanxing}. 

\section{Conclusion}\label{VIII}
In this paper, we have focused on enhancing the performance of blockchain-enabled Web 3.0, with a particular emphasis on optimizing block propagation in public blockchains. Specifically, we have introduced the AoB as a data-freshness metric to evaluate the efficiency of block propagation. Then, to ensure reliability during block propagation, we have established the miner reputation mechanism based on the subjective logic model, which can calculate miner reputations and prevent the new block from propagating to miners with lower reputations. Moreover, to achieve block propagation optimization, we have proposed a GAT-based reliable block propagation optimization model to minimize the overall AoB of block propagation with the given reputation constraint, thus obtaining the optimal block propagation trajectory. Finally, numerical results have demonstrated that compared with conventional routing mechanisms, the proposed scheme can minimize the overall AoB and ensure that miners have high reputation values during block propagation, contributing to enhanced efficiency and reliability of block propagation in public blockchains.

For future work, we will further explore whether advanced variants of GAT, such as sparse graph attention neural networks and graph sample and aggregate-attention networks, can achieve block propagation optimization more efficiently. In addition, we will deeply investigate the synergy between diffusion models and GAT to better capture the interaction between miners in complex environments, ultimately enhancing the efficiency and generalization ability of the model. Moreover, we will systematically explore how to expand the scale of graph miners to make the GAT model more applicable in real public blockchains.
\bibliographystyle{IEEEtran}
\bibliography{ref}
\end{CJK}
\end{document}